

\vsize=9.0in\voffset=1cm
\looseness=2


\message{fonts,}

\font\tenrm=cmr10
\font\ninerm=cmr9
\font\eightrm=cmr8
\font\teni=cmmi10
\font\ninei=cmmi9
\font\eighti=cmmi8
\font\ninesy=cmsy9
\font\tensy=cmsy10
\font\eightsy=cmsy8
\font\tenbf=cmbx10
\font\ninebf=cmbx9
\font\tentt=cmtt10
\font\ninett=cmtt9

\font\ninesl=cmsl9
\font\eightsl=cmsl8

\font\nineit=cmti9
\font\eightit=cmti8

\skewchar\ninei='177 \skewchar\eighti='177
\skewchar\ninesy='60 \skewchar\eightsy='60

\def\eightpoint{\def\rm{\fam0\eightrm} 
\normalbaselineskip=9pt
\normallineskiplimit=-1pt
\normallineskip=0pt

\textfont0=\eightrm \scriptfont0=\sevenrm \scriptscriptfont0=\fiverm
\textfont1=\ninei \scriptfont1=\seveni \scriptscriptfont1=\fivei
\textfont2=\ninesy \scriptfont2=\sevensy \scriptscriptfont2=\fivesy
\textfont3=\tenex \scriptfont3=\tenex \scriptscriptfont3=\tenex
\textfont\itfam=\eightit  \def\it{\fam\itfam\eightit} 
\textfont\slfam=\eightsl \def\sl{\fam\slfam\eightsl} 

\setbox\strutbox=\hbox{\vrule height6pt depth2pt width0pt}%
\normalbaselines \rm}

\def\ninepoint{\def\rm{\fam0\ninerm} 
\textfont0=\ninerm \scriptfont0=\sevenrm \scriptscriptfont0=\fiverm
\textfont1=\ninei \scriptfont1=\seveni \scriptscriptfont1=\fivei
\textfont2=\ninesy \scriptfont2=\sevensy \scriptscriptfont2=\fivesy
\textfont3=\tenex \scriptfont3=\tenex \scriptscriptfont3=\tenex
\textfont\itfam=\nineit  \def\it{\fam\itfam\nineit} 
\textfont\slfam=\ninesl \def\sl{\fam\slfam\ninesl} 
\textfont\bffam=\ninebf \scriptfont\bffam=\sevenbf
\scriptscriptfont\bffam=\fivebf \def\bf{\fam\bffam\ninebf} 
\textfont\ttfam=\ninett \def\tt{\fam\ttfam\ninett} 

\normalbaselineskip=11pt
\setbox\strutbox=\hbox{\vrule height8pt depth3pt width0pt}%
\let \smc=\sevenrm \let\big=\ninebig \normalbaselines
\parindent=1em
\rm}

\def\tenpoint{\def\rm{\fam0\tenrm} 
\textfont0=\tenrm \scriptfont0=\ninerm \scriptscriptfont0=\fiverm
\textfont1=\teni \scriptfont1=\seveni \scriptscriptfont1=\fivei
\textfont2=\tensy \scriptfont2=\sevensy \scriptscriptfont2=\fivesy
\textfont3=\tenex \scriptfont3=\tenex \scriptscriptfont3=\tenex
\textfont\itfam=\nineit  \def\it{\fam\itfam\nineit} 
\textfont\slfam=\ninesl \def\sl{\fam\slfam\ninesl} 
\textfont\bffam=\ninebf \scriptfont\bffam=\sevenbf
\scriptscriptfont\bffam=\fivebf \def\bf{\fam\bffam\tenbf} 
\textfont\ttfam=\tentt \def\tt{\fam\ttfam\tentt} 

\normalbaselineskip=11pt
\setbox\strutbox=\hbox{\vrule height8pt depth3pt width0pt}%
\let \smc=\sevenrm \let\big=\ninebig \normalbaselines
\parindent=1em
\rm}

\message{fin format jgr}
%
%
\magnification=1200
\font\toto=cmbx10 scaled 1200
\font\Bbb=msbm10
\def\N{\hbox{\Bbb N}}

\def\C{\hbox{\Bbb C}}
\def\R{\hbox{\Bbb R}}

\def\prc{(\hbox{\Bbb R}^2)}
\def\prb{C_{\infty}^{(\infty)}(\hbox{\Bbb R}^2)}
\def\r2{\hbox{\Bbb R}^2}
\let\\=\partial
\def\intt{\int\!\!\!\int}
\def\p{1\over (2\pi)^2}
\def\lp{\lambda^{\prime}}
\def\ls{\lambda^{\prime\prime}}
\def\l{\lambda}
\def\pl{\\_{\l}}
\def\pls{\\_{\bar\l}}
\def\a{\alpha}
\def\o{\bigl({1\over \lp}\bigr)^{2l+1}}
\def\s#1#2#3#4#5#6{\sum_{{#1}={#2}}^{{#3}}{#4}_{{#5}}({#6})}
\def\sg{-\pi\,{\rm sgn}\,(\l\bar\l-1)}
\def\ch#1#2#3{{\\{#1}_{{#2}}(\l,{#3},t)\over \\t}}
\tolerance=500
\noindent
{\toto Transparent Potentials at Fixed Energy in Dimension Two.}

\noindent
{\toto Fixed-Energy Dispersion Relations for the Fast Decaying Potentials.}
\vskip 8 mm
\noindent
{\bf Piotr\  G.Grinevich}$^{1,\star}$ \footnote{}{\ninerm $^\star$ The main
part  of this work was fulfilled during the visit of one of the authors
(P.G.G.) to the University of Nantes in June 1994. He is grateful to the
University of Nantes for the invitation and the financial support of this
visit. He was also supported  by the Soros International Scientific
foundation
grant MD 8000 and by the Russian Foundation for Fundamental Studies grant
93-011-16087.},
\ {\bf Roman\ G.Novikov}$^2$
\vskip 3 mm
\noindent
$ ^1\ $  {\ninerm  Landau\  Institute\ for\ Theoretical Physics,
 Kosygina 2, Moscow, 117940,Russia}

\noindent
{\ninerm e-mail:pgg@cpd.landau.free.net}

\noindent
$ ^2\ $ {\ninerm  CNRS, U.R.A. 758, D\'epartament de Math\'ematiques,
Universit\'e de Nantes,F 44072,Nantes,}

\noindent
{\ninerm \ \ \ \ cedex 03,France}

\noindent
{\ninerm e-mail:novikov@math.univ-nantes.fr}
\vskip 10 mm
\noindent
{\bf Abstract:}\  For the two-dimensional Schr\"odinger equation
$$ [- \Delta +v(x)]\psi=E\psi,\ x\in \R^2,\ E=E_{fixed}>0  \ \ \ \ \ (*)$$
at a fixed positive energy with a fast  decaying at infinity potential
 $v(x)$ dispersion relations on  the scattering data are given.Under
"small norm" assumption using these dispersion relations we give
(without a complete proof of sufficiency) a characterization of scattering
data for the potentials from the Schwartz class $S=C_{\infty}^{(\infty)}
(\hbox{\Bbb R}^2).$ For the potentials with zero scattering amplitude at a
fixed energy $\scriptstyle E_{fixed}$ (transparent potentials) we give a
complete proof of this characterization.
As a consequence we construct a family (parameterized  by a function
of one variable) of two-dimensional spherically-symmetric real potentials
from the Schwartz class $S$ transparent at a given energy.
For the two-dimensional case (without assumption that the potential is small)
we show that there are no nonzero real exponentially decreasing at infinity,
potentials transparent at a fixed energy.
For any dimension greater or equal 1 we prove that there are no nonzero real
potentials with zero forward scattering amplitude at an energy interval.
We show that KdV-type equations  in dimension 2+1 related with the scattering
problem $(*)$ (the Novikov-Veselov equations) do  not  preserve  , in general,
these dispersion relations starting from the second one.
As a corollary these equations do  not preserve,  in general , the decay rate
faster then $|x|^{-3}$ for initial data from  the Schwartz class.
\noindent
\vskip 6 mm
\noindent
{\bf Introduction.}

\noindent
An interesting property of the fixed-energy scattering  problem for the
Schr\"odinger equation in dimension 2
$$\displaystyle L\psi=E\psi,\ L=-{\partial^2\over {\partial x_1^2}}-
{\partial^2\over {\partial x_2^2}}+v(x),\ x=(x_1,x_2)\in \hbox{\Bbb R}^2,\
E\in \hbox{\Bbb R}\ (E\  {\rm is\  fixed}) \eqno (0.1)$$
is its deep connection with the soliton theory, i.e. the following methods
can be effectively applied to this problem: the finite-gap  technique, the
nonlocal Riemann problem method, the $\bar \partial$-problem method and this
problem possesses an infinite-dimensional algebra  of symmetries generated by
KdV-type equations in dimension 2+1 (Novikov-Veselov hierarchy). Scattering
transform for Eq. (0.1) allows us to integrate these equations. Inverse
scattering problem for (0.1) is closely connected also with the inverse
boundary value problem (Calderon problem).

The problems mentioned above were studied in the papers [1]-[22] and others
(some historical remarks are given in the end of this introduction). In the
present paper we study the scattering transform for the equation (0.1) for
potentials with decay rate at infinity $1/|x|^{M+2+\varepsilon},\
\varepsilon>0,\ M=0,1,2,\ldots.$ We show that such decay rate results in
M+1 algebraic relations on the scattering data (we shall call them
fixed-energy dispersion relations).

Let us recall the definition of the scattering data for (0.1).

We assume that
$$v(x)={\bar v}(x),\ v(x)\in L^{\infty}(\hbox{\Bbb R}^2),\
|v(x)|<q(1+|x|)^{-2-\varepsilon},\  \varepsilon>0,\ q>0, \eqno {(0.2)}$$
where $|x|=\sqrt{x_1^2+x_2^2}$.

For $E>0$ and any $k=(k_1,k_2)\in \hbox{\Bbb R}^2$, such that $k^2=E$ ,\
there exists an unique bounded solution $\varphi^{+}(x,k)$ of Eq. (0.1)
with the following asymptotics:
$$\displaystyle \varphi^{+}(x,k)=e^{ikx}-i\pi\sqrt{2\pi}e^{-i\pi\over 4}f
\bigl(k,|k|{x\over |x|}\bigr){e^{i|k||x|}\over \sqrt{|k||x|}}+o\bigl(
{1\over |x|}\bigr). \eqno (0.3)$$
The function $f(k,l)$ in (0.3),$ k\in \hbox{\Bbb R}^2,\ l\in
\hbox{\Bbb R}^2,\ k^2=l^2=E$ is called the scattering amplitude.

Let $k\in \hbox{\Bbb C}^2,\ k^2=E,\ {\rm Im}\ k\not =0$. Let, in addition
$\Delta(k)\not =0$, where $\Delta$ is the modified Fredholm determinant of
the integral equation (1.3). Then there exists an unique solution of (0.1)
such that
$$\psi(k,x)=e^{ikx}(1+ o(1)),\ {\rm  Im}\ k\not =0,\ {\rm for}\ |x|
\rightarrow \infty. \eqno (0.4)$$
It was shown in [12] that there exists a special real function $Q(|E|,\
varepsilon)$ with the following properties $Q(|E|,\varepsilon)>0$ as
$E\not =0,\ Q(|E|,\varepsilon)\rightarrow +\infty$ for fixed $\varepsilon$
as $|E|\rightarrow \infty$ such that if a potential $v(x)$ satisfies (0.2)
and
$$q<Q(|E|,\varepsilon ) \eqno  (0.5)$$
then

1)Fredholm determinant of the equation(1.3) $\Delta(k)\not =0$ for all
$k^2=E$.

2)The fixed-energy scattering data for the potential $v(x)$ is
"small enough" for unique solvability of the equations of inverse
scattering.

The "small norm" condition (0.5) means that the potential $v(x)$ is
small being compared with the energy.

The solutions of  the Schrodinger equation with asymptotics (0.5) were
introduced to the scattering theory by L.D.Faddeev [23] as solutions
of the integral equation (1.3). It can be shown [12] that for $E\in
\hbox{\Bbb R},\  {\rm Im}\ k\not =0$
$$\displaystyle \psi(k,x)=e^{ikx}-\pi {\rm sgn}({\rm Im}\ k_2{\bar k}_1)
e^{ikx}\bigl({a(k)\over {-k_2x_1+k_1x_2}}+{e^{-2i{\rm Re} k\cdot x}\over
{-{\bar k}_2x_1+{\bar  k}_1x_2}}b(k)+o\bigl({1\over |x|}\bigr)\bigr),
\eqno (0.6)$$
where the function $a(k)$ and $b(k)$ are expressed through the Faddeev's
scattering data  by the formula (1.7). The formula similar to (0.6) can
be written for any complex E.

We consider  the functions $a(k)$ and $b(k)$ as  additional scattering
data to $f(k,l)$ for $E>0$ and as the main scattering data for other $E$.

Using the results of [8,\ 9] it was shown in [12] that at fixed energy
under conditions (0.2),(0.5) the scattering amplitude $f(k,l)$ and the
function $b(k)$ uniquely determine the potential. From the inverse
scattering problem it follows that in the slow decaying case $f(k,l)$
and $b(k)$ are independent at fixed energy.

For the potentials exponentially decreasing at infinity the uniqueness
of the reconstruction via  the fixed energy scattering amplitude was
proved in [9,\ 12] for the two-dimensional case under the conditions
(0.2),(0.5) at the fixed energy and for the three-dimensional case in
[24] with and in [27] without the "small norm" assumption.

The scattering amplitude at a fixed energy is insufficient, in general,
to reconstruct the potential  uniquely.

In the exact formulation it was shown in the series of papers
[31],[32],[33] and others started by fundamental paper of T.Regge [31].
In the works of this series the fixed-energy inverse scattering probllem
was studied in the 3-dimensional spherically-symmetrical case. The
existence of nonzero multidimensional potentials with zero scattering
amplitude at a fixed energy (transparent at a fixed energy potentials)
was shown by R.G.Newton [32]. The properties of these potentials were
clarified by P.C.Sabatier in [33], where one-dimensional family of
transparent at a fixed energy potentials was given and it was shown
that nonzero potentials from this family decrease at  infinity as
$|x|^{-3/2}$.

In [8] it was shown that transparent at a fixed energy two-dimensional
potentials with the "small norm" assumption are parameterized by a
function of two variables. From the results of the present paper  it
follows that constructed in [8] transparent potentials decrease  , in
general, as $|x|^{-2}$. Explicit real nonsingular rational two-dimensional
potentials with zero scattering amplitude at a fixed energy were given in
[10]. They also decrease as $|x|^{-2}$.

The central point of the present paper is a characterization of the
scattering data at fixed positive energy for the real-valued potentials
of the Schwartz  class $S=C_{\infty}^{(\infty)}(\hbox{\Bbb R}^2)$.
On the basis of this characterization we  construct (Proposition 1,\
Theorem 2) real two-dimensional spherically-symmetric potentials from
the Schwartz class $S$ with zero scattering amplitude at a fixed energy
$E>0$. The classical scattering solution $\varphi^{+}(x,k)$ for such
potentials has the following asymptotics at infinity
$$\varphi^{+}(x,k)=e^{ikx}+O\bigl(1/|x|^{\infty}\bigr)\ {\rm for}\ k^2=E.
\eqno (0.7)$$
Further, (Theorem 3) we prove the following statement . Let the fixed
energy scattering amplitudes of two exponentially decreasing potentials
with the property (0.2) coincide and one of these potentials possesses ,
in addition, the property  (0.5) at this fixed energy. Then these two
potentials coincide. This statement improves the corresponding theorem
from [9,\ 12]. In particular, there exists no nonzero two-dimensional
exponentially decreasing real nonsingular potentials transparent  at a
fixed energy.

We prove that there are no nonzero real potentials transparent at an
energy interval. Moreover,  we prove that if the forward scattering
amplitude is equal to zero at an energy interval then the real potential
is equal to zero identically (Theorem 4). This result is valid without
the small norm assumption in any dimension greater or equal to 1.

The most nontrivial part of our characterization theorem is the existence
of additional algebraic relations on the scattering data\ --\ fixed
energy dispersion relations.

Let the potential $v(x)$ satisfy (0.2),(0.5). Then the scattering
amplitude $f(k,l)$ satisfies (1.29) and $b$ satisfies (1.25). Assume
now, that, in addition, the potential $v(x)$ belongs to the Schwartz
class $S=C_{\infty}^{(\infty)}(\hbox{\Bbb R})$. Then for functions $f,b$
we have (3.8),(3.9). In the inverse problem we may start from arbitrary
functions $f,b$ satisfying (1.29),(3.8) and (1.25),(3.9) respectively
which are "small" enough for unique solvability of integral equations
of the inverse problem but the corresponding potential may decrease at
infinity rather slow. Necessary conditions on the scattering data for
the fast decaying potentials were found in [9,\ 12] for the positive
energy case. Another  set of necessary conditions for the fast  decay
rate  were  found in [11] for the negative energy case. In the present
paper we show that analogs of the necessary conditions from [11] (we
call them fixed-energy dispersion relations) are valid in the positive
case too.
(For three-dimensional problem an analog of the first dispersion relation
was used in [24]).

In the present paper we show that for real potentials from the class
$S$ under  the "small norm" assumption
the scattering data $f(k,l),b(k)$ satisfy (1.29),(3.8),(1.25),(3.9) and
$2\cdot \infty +2$ additional conditions from the section 3 corresponding
to $M=\infty$ are fulfilled. Let $f(k,l),b(k)$ be  arbitrary functions
satisfying
(1.29),(3.8),(1.25),(3.9) and $2\cdot \infty +2$ additional conditions
from section 3 . Assume also that $f(k,l),b(k)$ are sufficiently small, so
the integral equation of the inverse problem  has unique solution. Then our
hypothesis is that the corresponding potential is from the class $S$.
Some restriction in time give  us no possibility to carry out in the
present paper a complete proof of this hypothesis.  In the present paper
we prove this hypothesis (Theorem 1) in the transparent case
$f(k,l)\equiv 0,\ k^2=l^2=E$ at fixed energy E.  In  this case the
"small norm" assumption for $b(k)$ is not necessary.

Results on inverse scattering at  fixed   energy for Eq. (0.1) can   be
 applied to the solution    of the Cauchy problem (and to the construction
of explicit soliton type solutions) for the KdV-type equation in dimension
$2+1$ (Novikov-Veselov equation)
$$\eqalignno{
{\partial v(x_1,x_2,t)\over \partial t}&=2{\partial^3v \over \partial x^3}
-6{\partial^3v\over {\partial  x_1  \partial x_2}}+2\bigl({\partial u\over
  \partial x_1}+{\partial w\over \partial x_2}\bigr)-2E\bigl({\partial u
\over  \partial x_1}+{\partial w\over \partial x_2}\bigr) \cr
v&=\bar v,\ E\in \hbox{\Bbb R},\ x,y,z\in \hbox{\Bbb R} \cr\cr
 u(x_1,x_2,t)&={3\over \pi}\intt_{\R^2}{{v(x_1^
{\prime},x_2^{\prime},t)((x_1-x_1^{\prime})^2-(x_2-x_2^{\prime})^2)}\over
{((x_1-x_1^{\prime})^2+(x_2-x_2^{\prime})^2)^2}}dx_1^{\prime}dx_2^{\prime}
&(0.8)\cr\cr
w(x_1,x_2,t)&=-{6\over \pi}\intt_{\R^2}{{v(x_1^{\prime}
,x_2^{\prime},t)(x_1-x_1^{\prime})(x_2-x_2^{\prime})}\over
{((x_1-x_1^{\prime}
)^2+(x_2-x_2^{\prime})^2)^2}}dx_1^{\prime}dx_2^{\prime} \cr}$$
and its higher analogs. The equation (0.8) is contained implicitly in the
paper of S.V.Manakov [1] as an  equation possessing the following
representation
$${\partial (L-E)\over \partial t}=[L-E,A]+B(L-E), \eqno (0.9) $$
(Manakov $L-A-B$ triple), where $L$ is the   Schr\"odinger operator from
(0.1),
$A$ and $B$ are suitable differential  operators   of the third and zero
order respectively. The equation (0.8) was written in an explicit form by
S.P.Novikov and A.P.Veselov in [3,\ 4], where higher analogs of (0.8) were
also constructed.

The both Kadomtsev-Petviashvily equations can  be obtained from (0.8) by
considering an appropriate limit $E\to\pm\infty$
 (V.E.Zakharov, private communication).
In terms of the scattering data  the nonlinear equation (0.8) takes the
form
$${\eqalign{
{\partial b(k)\over \partial t}&=2i[k_1^3+{\bar k}_1^3-3k_1k_2^2-
3{\bar k}_1{\bar k}_2^2]b(k),\ k\in \C^2,\ {\rm Im}\ k\ne 0,\ k^2=E, \cr
{\partial f(k,l)\over \partial t}&=2i[k_1^3-3k_1k_2^2-l_1^3+3l_1l_2^2]
f(k,l),\ k,l\in \R^2,\ k^2=l^2=E. \cr}} \eqno (0.10)$$
In the present paper (Corollary 1,Theorem 1,Theorem 5) we obtain the
following result.

Let $v(x,t)$ be a solution of (0.8) with the following Cauchy data
$v(x)=v(x,0):$

1)\ $v(x)\in C_{\infty}^{(\infty)}(\hbox{\Bbb R}^2)$,

2)\ $v(x)$ satisfies (0.5),

3)\ $v(x)$ is transparent at the energy $E$\ i.e. $f(k,l)\equiv 0$ at
the energy E,

4)\ $v(x)\not\equiv 0.$

Then for any $t\ne
0\ \ v(x,t)\in C_3^{(\infty)}\prc\ {\rm and} \ v(x,t)\not\in
C_{3+\varepsilon}^{(0)}\prc$ \ (i.e. $v(x,t)$ decreases at $|x|\to
\infty$ exactly as $|x|^{-3}$).

In the theorem 5 under the "small norm" assumption we  obtain,
in particular, the following result. Let the Cauchy data $v(x,0)\in
C_{\infty}^{(\infty)}\prc\  $ generate a solution $v(x,t)\   $ of
(0.8) such that at a fixed $t\not =0 \ \ v(x,t)$ decreases at infinity
as $|x|^{-3-\varepsilon},\ \varepsilon>0$. Then
$$\displaystyle \int_{x\in \hbox{\Bbb R}^2}v(x)dx=0. \eqno (0.11)$$
We have, also, the following hypothesis. Under "small norm" assumption
the Cauchy data $v(x,0)\in \prb\ $ for the equation (0.8) generates
a solution
$v(x,t)\in C_{2+\varepsilon}^{(\infty)}\prc\ \ {\rm in}\ \ x,$

$ 0<\varepsilon < {1\over 2}$. This solution belongs to
$C_{3+\varepsilon}^{(\infty)}\prc\ {\rm in}\ x $ if (0.11) is fulfilled.
The faster decay rate for all $t$ results in additional conditions on
the Cauchy data  which can be written. We think, also, that this
hypothesis is true without the "small norm" assumption, but it is not
clear for us how to prove it in the latter case.

Let us mention the following. The decay rate of the potentials constructed
in the preceding papers was not studied carefully enough.  For example,
we correct corollary 1 from the paper [8] and proposition 9.4 from [12].
\bigskip
\noindent
{\bf Historical remarks.}
\medskip
\noindent
The relations between the fixed-energy scattering transform for the
two-dimensional Schr\"odinger operator and nonlinear integrable
equations in dimension $2+1$ were observed for the first time by
S.V.Manakov [1].

The methods of the soliton theory were applied for the first time to  t
he inverse problem at fixed energy for the two-dimensional Schr\"odinger
operator in 1976 by B.A.Dubrovin, I.M.Krichever,S.P.Novikov [2] in the
quasiperiodic case.

The sufficient conditions on the finite-gap scattering data which
guarantee absence of the magnetic field and reality of the potential
were found by S.P.Novikov and A.P.Veselov in [3,\ 4].

The nonlocal Riemann problem method (Manakov [28]) together with ideas
from [3,\ 4] were applied by the authors in [5,\ 6]  for constructing
two-dimensional Schr\"odinger operators with decreasing potentials and
explicit solutions of the corresponding KdV-type equations in dimension
$2+1$ .

The connection between the kernel of the nonlocal Riemann problem and
the scattering amplitude at the fixed energy for the corresponding
potentials was found by one of the authors (R.G.N.) in [7]. As a
consequence a characterization of the scattering amplitude at a fixed
energy with small norm for real, smooth, decaying at infinity potentials
was obtained in [7].

The scattering transform at a fixed energy for general decaying at
infinity two-dimensional potentials was constructed by Manakov and
one of the authors (P.G.G.) in [8]. In [8] it was shown that in this
scattering transform the nonlocal Riemann problem of the type  [28] and
the $\bar\partial$-problem  of  the type [29] are present simultaneously.
In [8] it was shown that the connection between the fixed-energy scattering
amplitude and the nonlocal Riemann problem data found in [7] is unchanged in
the presence of nontrivial $\bar\partial$-problem data.
Thus, the $\bar\partial$-problem data parameterizes the variety of all
potentials with the given fixed-energy scattering amplitude. Assuming the
nonlocal Riemann problem data to be identically zero transparent  at a
fixed energy potentials were  obtained in [8]. The "spectral transform"
constructed in [8] was applied to solve the Cauchy problem for equations
from the KdV-type Novikov-Veselov hierarchy for the decaying at infinity
Cauchy data.

In [9] it was shown by one of the authors (R.G.N.) that the scattering
data introduced in
 [8] can be considered as a restriction of the Faddeev scattering data
 [23] to a fixed energy level.The connection between the scattering
 amplitude at a fixed energy and the nonlocal Riemann problem data was
 obtained in [9] ones again from the point of view of the direct problem
 by the technique developed in [23]. In [9] the necessary conditions on
 the fixed energy scattering data corresponding to the fast decaying at
 the infinity potentials were found and it was shown that exponentially
 decreasing potentials  under small norm assumption are uniquely determined
 by the fixed-energy scattering amplitude.

The explicit examples of real nonsingular transparent  at a fixed energy
potentials (rational solitons) were constructed by  one of the authors
(P.G.G.) in [10] .
These potentials decay at infinity rather slow (as the minus second power
of distance). These potentials are constructed independently by
V.E.Zakharov.

The two-dimensional scattering problem at a fixed  negative energy was
studied by S.P.Novikov and by  one of the authors (P.G.G.) in [11]. In
this case we have a pure $\bar\partial$-problem. In [11] it was shown
that for an arbitrary nonsingular scattering data (without the small
norm assumption) satisfying the reality and the absence of magnetic
field reduction the solution of the inverse problem is unique and
nonsingular and the $L^2$ spectrum of the corresponding operator lies
above our fixed energy. If it is not so the $\bar\partial$-problem data
is singular but rather little about inverse scattering in this case is
known.
For the fast decaying at infinity potentials the necessary conditions
on the scattering data were found in [11]. These conditions have a
different origin and a different structure than the necessary  conditions
from [9].

The further developments and generalization of these papers [5-11] and
some results of [23-26,\ 18,\ 30] were  given in [12].

In [13] by J.-P.Fransoise and one of the authors (R.G.N.) the hamiltonian
systems describing dynamics of poles  of the rational solitons from [10]
were found.

In papers [5-12] only the case of nonzero fixed energy was studied.
The zero energy level was examined by M.Boiti,J.Leon,M.Manna,F.Pempinelli
[14],T.Y.Tsai [15],Z.Sun,
\noindent
G.Uhlmann [21],A.Nachman [22].

On the other hand the studies of the inverse problem at fixed energy
$(E=0)$ for the two-dimensional Schr\"odinger equation (for the equation
${\rm div}(\gamma(x){\rm grad} \psi)=0$) in a bounded domain were
stimulated by the paper [16] of A.P.Calderon.   In the two-dimensional
case the studies of the Calderon problem were  started by
R.Kohn,M.Vogelis [17], J.Sylvester and G.Uhlmann [18].
The method to apply results of the two-dimensional inverse scattering
at fixed energy to the Calderon problem was given  for the first time
by one of the authors  (R.G.N.) in [19]. Among subsequent works on
the Calderon problem in dimension 2 let us mention important papers
of Z.Sun,G.Uhlmann [20,21] and A.Nachman [22].

\medskip
\noindent
{\bf 1.The equations of direct scattering.}
\medskip
\noindent
The Faddeev scattering data (see [23,\ 24]) $h(k,l),\ k,l\in \hbox{\Bbb C}^2,
\ k^2=l^2=E,\ {\rm Im}\ k =\ {\rm Im}\ l\ $ for equation (0.1) are defined
by the formula
$$\displaystyle h(k,l)={1\over (2\pi)^2}\intt_{\R^2} e^{-ilx}\psi(x,k)
v(x)dx_1dx_2,  \eqno (1.1)$$
where
$$\psi(x,k)=e^{ikx}\mu(x,k),  \eqno (1.2)$$
$$\displaystyle \mu(x,k)=1+\int\!\!\!\int_{\hbox{\Bbb R}^2} g(x-y,k)v(y)
\mu(y,k)dy_1dy_2, \eqno (1.3)$$
$$\displaystyle g(x,k)=-{1\over (2\pi)^2}\intt_{\R^2} {e^{i\xi x}\over {
\xi^2+2k\xi}}d\xi_1d\xi_2,\ \ {\rm Im}\ k \not =0. \eqno (1.4)$$
For $k\in \r2\ $ the following limits exist:
$$\psi_{\gamma}(x,k)=\psi(x,k+i0\gamma),\ \mu_{\gamma}(x,k)=\mu(x,
k+i0\gamma),$$
$$h_{\gamma}(k,l)=h(k+i0\gamma,l+i0\gamma),\ k,l,\gamma\in \r2\ ,k^2=
l^2=E,\ \gamma^2=1.  \eqno (1.5)$$
In addition,
$$\varphi^{+}(x,k)=\psi_{k/|k|}(x,k),\ f(k,l)=h_{k/|k|}(k,l), \eqno (1.6)$$
where $\varphi^{+},\ f$\ are functions from (0.3). For $k^2=E\in
\hbox{\Bbb R}
,\ {\rm Im}\ k \ \not =0,\ \ \psi(x,k)$ is the  function (0.6). For $a(k)$
and
 $b(k)$ the following formulas are valid
$$a(k)=h(k,k),\ b(k)=h(k,k+\xi (k)), \eqno (1.7)$$
where $\xi (k)\ $ is a different from zero root of the equation
$$\xi^2\ +\ 2k\xi\ =\ 0,\ \ \xi\in \r2. \eqno (1.8)$$
In the two-dimensional fixed-energy scattering theory it is convenient to
 introduce    new notations
$${\eqalign{
z=x_1+ix_2,\ \bar z=x_1-ix_2,\ \\_z={1\over 2}(\\_x-i\\_y),\
\\_{\bar z}={1\over 2}(\\_x+i\\_y), \cr
\lambda={{k_1+ik_2}\over \sqrt{E}},\ \lambda^{\prime}={{l_1+il_2}\over \sqrt
{E}},\ E=k_1^2+k_2^2=l_1^2+l_2^2.\cr}} \eqno (1.9)$$
In addition,
$$k_1={\sqrt{E}\over 2}\bigl(\lambda+{1\over \lambda}\bigr),\ k_2={i\sqrt{E}
\over 2}\bigl({1\over  \lambda}-\lambda\bigr),\ e^{ikx}=e^{{i\over 2}\sqrt{E}
(\lambda \bar z+z/\lambda)}. \eqno (1.10)$$
In new notations the Schr\"odinger equation (0.1) takes the form
$$L\psi=E\psi,\ L=-4\\_z\\_{\bar z}+v(z),\ z\in \hbox{\Bbb C}^1,\ E\in
\hbox{\Bbb R} \eqno (1.11)$$
(in this paper the notation $f=f(z)$ does not mean that $\\_{\bar z}f=0$).

The functions $\varphi^{+}$ from (0.3),\ $\psi,\ \mu\ $ from (1.2),(1.3),
\ $a,b$ from (1.7) take the form
$${\eqalign{
\varphi^{+}=\varphi^{+}(z,\lambda,E),\ f=f(\lambda,\lambda^{\prime},E),\
\psi=\psi(z,\lambda,E),\ \cr
 \mu=\mu(z,\lambda,E),\ a=a(\lambda,E),
b=b(\lambda,E). \cr}}
 \eqno (1.12)$$
Further, we shall always assume that the  fixed energy
$$E=1 \eqno (1.13)$$
(the case of an arbitrary fixed positive energy may be  reduced to (1.13)
by scaling transformation). We shall also omit $E$ in the further notations.

Now (1.2) reads as
$$\psi(z,\lambda)=e^{{i\over 2}(\lambda \bar z +z/\lambda)} \mu(z,\lambda),
\eqno (1.14)$$
for the functions $a(\lambda),\ b(\lambda)$ we can write
$${\eqalign{
a(\lambda)&= {\p} \  \intt_{\C} v(z)\mu(z,\lambda)dz_R dz_I, \cr
b(\lambda)&={\p} \ \intt_{\C} e^{{i\over 2}(\lambda \bar z+\bar
\lambda z+z/\lambda+
\bar z/\bar\lambda)}v(z)\mu(z,\lambda)dz_Rdz_I.\cr}} \eqno (1.15)$$
For $|\lambda|=1$ corresponding to ${\rm Im}\ k=0$ formulas (1.3),(1.4)
make no sense without a regularization, but the boundary values
$$\mu_{\pm}(z,\lambda)=\mu(z,\lambda (1\mp 0)), \eqno  (1.17)$$
are well-defined. We consider also functions
$$\displaystyle h_{\pm}(\lambda,\lambda^{\prime})={\p} \ \intt_{\C}
e^{-{i\over 2}(\lambda^{\prime}\bar z+z/\lambda^{\prime})}v(z)\Psi_{\pm}(z,
\lambda)dz_Rdz_I, \eqno (1.18)$$
where
$$|\lambda|=|\lambda^{\prime}|=1,\ h_{\pm}(\lambda,\lambda^{\prime})=
h_{\pm n_{\bot}}(k,l),\ n_{\bot}=(-k_2,k_1)/|k|.$$
Let the potential $v(z)$ satisfy the "small norm" condition (0.5) for
$E=1$. Then the function $\psi(z,\lambda)$ has the following properties
(see [8,\ 12]):

1)\ For all $|\lambda|\ne  1\ \ \psi(z,\lambda)$ is uniquely defined by
the equation (1.3).

2)\ $\psi(z,\lambda)$\ is continuous in $\lambda$ outside the unite
circle $|\lambda|=1$.

3)\ There exists a function $\rho(\l,\lp),\ |\l|=|\lp|=1\ $ such that the
boundary values of the function $\psi(\l,z)$\ on the unit circle $|\l|=1$\
satisfy
$$\displaystyle \psi_{+}(z,\l)=\psi_{-}(z,\l)+\oint_{|\lp|=1}\rho(\l,\lp)
\psi_{-}(z,\lp)|d\lp|. \eqno (1.19)$$
4)\ Outside the  unit circle the function $\psi(z,\l)$\ satisfies  the
following equation
$${\\ \psi(z,\l)\over \\ {\bar \l}}=r(\l)\psi(z,-1/\bar \l), \eqno (1.20)$$
where
$$\displaystyle r(\l)={\pi {{\rm sgn}(\l\bar\l-1)}\over \bar\l}b(\l).
\eqno (1.21)$$
In terms of $\mu(z,\l)$ the equations (1.19) and (1.20) take the form
$$\displaystyle \mu_{+}(z,\l)=\mu_{-}(z,\l)+\oint_{|\lp|=1}\rho(\l,\lp,z)
\mu_{-}(z,\lp)|d\lp|, \eqno (1.19^{\prime})$$
$${\\ \mu(z,\l)\over \\ \bar\l}=r(\l,z)\mu(z,-1/\bar\l), \eqno (1.20^{
\prime})$$
where
$${\eqalign{
\rho(\l,\lp,z)&=e^{-{i\over 2}(\l\bar z+z/\l-\lp \bar z-z/\l)}
\rho(\l,\lp),\cr
r(\l,z)&=e^{-{i\over 2}(\l \bar z+z/\l+\bar \l z +\bar z/\bar\l)}r(\l). \cr}}
\eqno (1.22)   $$
5)
$$\psi(z,\l)=e^{{i\over 2}(\l \bar z +z/\l)}(1+o(1))\  {\rm as}\
\l\to 0,\infty,
\eqno (1.23)$$
$$\mu(z,\l)\to 1 \ {\rm as}\ \l\to 0,\infty. \eqno (1.23^{\prime}) $$
The functions $\rho(\l,\lp),\ b(\l),\ \psi(z,\l)$ has the following symmetry
properties [5,\ 6,\ 8]:
$$\rho(\l,\lp)+\rho(-\lp,-\l)+\oint_{|\lp|=1}\rho(\l,\lp)\rho(-\ls,-\lp)
|d\lp|=0
\eqno (1.24a)$$
for all $\l,\lp,\ |\l|=|\lp|=1,$
$$\rho(\lp,\l)=\overline{\rho(\l,\lp)},  \eqno (1.24b)$$
$$b(1/\bar\l)=b(\l), \eqno (1.25a)$$
$$b(-\l)=\overline{ b(\l)}, \eqno (1.25b)$$
$$\psi(z,-1/\bar\l)=\overline{\psi(z,\l)},\ \mu(z,-1/\bar\l)=
\overline{\mu(z,\l)}. \eqno (1.26)$$
Using (1.25) we may rewrite equations $(1.20),(1.20^{\prime})$ as
$${\\ \psi(z,\l)\over \\ \bar\l}=r(\l)\overline{\psi(z,\l)},\  \
{\\ \mu(z,\l)\over \\ \bar\l}=r(\l,z)\overline{\mu(z,\l)}. \eqno (1.27)$$
The scattering amplitude $f(\l,\lp)$  and the function $\rho(\l,\lp)$ are
connected with $h_{\pm}(\l,\lp)$ by the  following equations (see [12]):
$$\displaystyle h_{\pm}(\l,\lp)-\pi i \oint_{|\ls|=1}h_{\pm}(\l,\ls)\theta
\big[\pm {1\over i}\bigl({\ls\over \l}-{\l\over  \ls}\bigr)\big]
f(\ls,\lp)|d\ls|=f(\l,\lp) \eqno (1.28a) $$
$$b_{\pm}(\l,\lp)=\theta\big[-{1\over i}\bigl({\lp\over \l}-
{\l\over\lp}\bigr)
\big]h_{\pm}(\l,\lp)-\theta\big[{1\over i}\bigl({\lp\over \l}-
{\l\over \lp}
\bigr)\big]h_{\mp}(\l,\lp) \eqno (1.28b)$$
$$\rho(\l,\lp)+\pi  i \oint_{|\ls|=1}\rho(\l,\ls)\theta\big[\pm {1\over i}
\bigl({\lp\over \ls}-{\ls\over \lp}\bigr)\big]h_{\pm}(\ls,\lp)|d\ls|=-\pi i
h_{\pm}(\l,\lp) \eqno (1.28c)$$
(here $\theta(x)$ is the standard Heaviside function $\theta(x)=0,\ x<0,\
\theta(x)=1,\ x\ge  0$).

It is well known that for a real sufficiently fast decreasing at infinity
potential the scattering amplitude has the following properties (see,for
example,  [35]):

a)\ Reciprocity
$$f(-\lp,-\l)=f(\l,\lp). \eqno (1.29a)$$

b)\ Unitarity
$$\displaystyle f(\l,\lp)-\overline{ f(\lp,\l)}+\pi i
\oint_{|\ls|=1}f(\l,\ls)
 \overline{ f(\lp,\ls)}|d\ls|=0. \eqno (1.29b)$$
Due to the equations (1.28) the property (1.29j) implies (1.24j), where
$j=a,b$ and vice versa [7,\ 12]. In  terms of $h_{+}(\l,\lp),\ h_{-}(\l,\lp)$
defined by (1.28a) the properties (1.29) take the form (see [12])
$${\eqalign{
h_{-}(\l,\lp)&=h_{-}(-\lp,-\l)+\pi i \oint h_{-}(\l,\ls)h_{-}(-\lp,-\ls) \cr
&\times\big[\theta\bigl(-{1\over i}\bigl({\ls\over \l}-{\l\over \ls}\bigr)
\bigr)-
\theta\bigl(-{1\over i}\bigl({\ls\over \lp}-{\lp\over\ls}\bigr)\bigr)\big]
|d\ls|, \cr
&h_{+}(\l,\lp)=\overline {h_{-}(-\lp,-\l)},\cr\cr}} \eqno (1.30)$$
where (1.28a) is also assumed to be valid.

It is well-known also that under condition (0.2) the scattering amplitude
$f(\l,\lp)$ is a continuous function . If (0.2) and (0.5) are valid then
$b(\l)$ is continuous for $|\l|\ne 1,$ and
$$r(\l)={\pi\over \bar\l}{\rm sgn}\ (\l\bar\l-1)b(\l)\in L_{p,2}(\hbox
{\Bbb C}), \eqno (1.31)$$
where $2<p<4 $ (see [12]).
\medskip
\noindent
{\bf 2.The equations of the inverse scattering.}
\medskip
\noindent
Given scattering data at fixed energy $E=1 \ f $\ and $b$,
where $f=f(\l,\lp),
|\l|=|\lp|=1$ is an arbitrary continuous function satisfying (1.24)
and $b(\l)$
is an arbitrary continuous function in the domains
$D_{\pm}=\{\l\in \hbox{\Bbb C}\ | \pm |\l|\le \pm 1\}$
satisfying (1.25) such that
$$|b(\l)|\le {b_{0}\over (|\l|+|1/\l|)^3}.\eqno(2.1) $$
(The boundary values of $b(\l)$ on the unit circle $|\l|=1$ in $D_{+}$ and
$D_{-}$\  may be different.) Then the corresponding potential $v(z)$ is
 constructed    in     the following way (see [8,\ 12]).

1)\ Using equations (1.28) we calculate $\rho(\l,\lp)$ via $f(\l,\lp)$ and
define $r(\l)$ by (1.21).

2)\ We construct a function $\mu(z,\l)$ with the analytic  properties
$(1.19^{\prime}),(1.20^{\prime}),(1.23^{\prime})$ as a solution of the
following integral equation
$$\displaystyle {\eqalign{
\mu(z,\l)=1&+{1\over {2\pi i}}\oint_{|\zeta|=1}{d\zeta\over {\zeta-\l}}
\oint_{|\lp|=1}\rho(\zeta,\lp,z)\mu(z,\lp(1-0))|d\lp| \cr
&-{1\over \pi}\intt_{\hbox{\Bbb C}}r(\zeta,z)\overline{\mu(z,\zeta)}
{d\zeta_Rd\zeta_I\over {\zeta-\l}}. \cr}} \eqno (2.2)$$
Here the Cauchy-Green formula was used
$$\displaystyle
f(\l)=-{1\over \pi}\intt_{D}(\\_{\bar\zeta}f(\zeta)){d\zeta_Rd\zeta_I\over
{\zeta-\l}}+{1\over {2\pi i}}\oint_{\\ D}f(\zeta){d\zeta\over {\zeta-\l}}.
\eqno (2.3) $$
The main case of our paper is $f(\l,\lp)\equiv 0$. From (1.28) it follows
that $\rho(\l,\lp)\equiv 0$. In this   case the equation (2.2) is uniquely
solvable in $C(\hbox{\Bbb C})$ for all $z$ under condition (2.1)
on the scattering data.

For the case of negative energy this fact was used in the paper [11] and
then for the case of positive energy in [12]. Another system of integral
equations for solving $(1.19^{\prime}), (1.20^{\prime}),(1.23)$, which
is more convenient in the case $f(\l,\lp)\ne 0$ was suggested in [12].

3)\ Expanding $\mu(z,\l)$ as $\l\to \infty$,
$$\mu(z,\l)=1+{\mu_{-1}(z)\over \l}+o\bigl({1\over \l}\bigr)  \eqno (2.4)$$
(from (2.1) it follows that there is no $c(z)/\bar\l$ term in (2.4)) we
define $v(z)$ by the formula
$$v(z)=2i\\_z \mu_{-1}(z). \eqno (2.5) $$
4)\ It can be shown (see [8]) that
$$L\psi(z,\l)=\psi(z,\l), \eqno (2.6)$$
where
$$\psi(z,\l)=e^{{i\over 2}(\l\bar z+z/\l)}\mu(z,\l),\ L=-4\\_z \\_{\bar z} +
v(z), \eqno (2.7)$$
$$\overline{v(z)}=v(z). \eqno (2.8)$$
Potential $v(x)$ constructed from the scattering data $f(\l,\lp),b(\l)$
with properties formulated in the beginning of this section may  decay rather
slowly. The necessary and sufficient conditions for decay rate at infinity
faster than $|x|^{-M},\ M>0$ will be discussed in the next sections.
\bigskip
\noindent
{\bf 3.Fast decaying potentials. Necessary conditions on the scattering data.}
\medskip
\noindent
Later we  shall use the following notation
$$\displaystyle {\eqalign{
v(x)\in C_M^{(N)}\prc\ \ {\rm if}\  {\\^{n_1+n_2}\over {\\ x_1^{n_1} \\ x_2^
{n_2}}}v(x)\in C\prc\ \  {\rm and} \cr
\bigg|{{\\^{n_1+n_2}}\over {\\x_1^{n_1}\\x_2^{n_2}}}v(x)\bigg|<{c_{n_1,n_2}
\over (1+|x|)^M},\  \ c_{n_1,n_2}>0 \cr}}  \eqno (3.1)$$
for all nonnegative integers $n_1,n_2$ such that $n_1+n_2\le N$.

Let the potential $v(x)$ satisfy (0.2),"small norm" assumption (0.5) and
$$v(x)\in C_{M+2+\varepsilon}^{(3)}\prc\ . \eqno (3.2)$$
In  this section we show that  under these assumptions we  have $2M+2$
 additional necessary conditions on the scattering data.

Let us introduce the  following functions:
$$\displaystyle a_m(\l)={\p}\intt_{\C}\bigl[\bigl({\\ \over
{\\ \l}}\bigr)^m e^
{-{i\over 2}[\l\bar z+z/\l]}\bigr]v(z)\psi(z,\l)dz_Rdz_I, \eqno (3.3a)$$
$$\displaystyle b_m(\l)={\p}\intt_{\C} \bigl[\bigl({\\ \over
{\\ \bar\l}}\bigr)^m
e^{{i\over 2}[\bar\l z+\bar z/\bar \l]}\bigr]v(z)\psi(z,\l)dz_Rdz_I.
\eqno (3.3b)$$
(It should be  noted that $a(\l)=a_0(\l),\ b(\l)=b_0(\l)$).

If (0.2),(0.5),(3.2) are fulfilled then

1)\ For $m=0,1,\ldots,M$ the  integrals (3.3) converge, the functions
$a_m(\l),b_m(\l)$ are  continuous in $D_{-}$ and $D_{+}\backslash 0$.

2)\
$$\eqalignno{
{\rm as}\ |\l|\to \infty \ \  &a_m(\l)=O(1),  &(3.4a-) \cr
		 &b_m(\l)=O\bigl({1\over |\l|^3}\bigr).  &(3.4b-)\cr
{\rm as}\ |\l|\to 0 \ \ &a_m(\l)=O\bigl({1\over |\l|^{2m}}\bigr), &(3.4a+)
\cr
     &b_m(\l)=O\bigl({|\l|^3\over |\l|^{2m}}\bigr). &(3.4b+)\cr}$$
{}From  (1.20) and (3.3) it follows that

3)
$$\eqalignno{
\\_{\bar\l}a_m(\l)&=r(\l)\overline{ b_m(\l)}, &(3.5a)\cr
\\_{\bar\l}b_m(\l)&=b_{m+1}(\l)+r(\l)\overline{  a_m(\l)}, &(3.5b)\cr}$$
where $r(\l)=\pi {\rm sgn} (\l\bar\l-1)b(\l)/\bar\l.$

In the formulas (3.3) we apply the operators $\\_{\l}$ and $\\_{\bar\l}$
to a holomorphic function and to an antiholomorphic function respectively.
Let
$$\l=re^{i\varphi},\ \lp=r^{\prime}e^{i\varphi^{\prime}},\ r,r^{\prime}\in
\hbox{\Bbb R}_{+}.$$
Then
$$\\_{\l}={\bar\l\over \l}\\_{\bar\l}+{1\over {i\l}}\\_{\varphi},\
\\_{\bar\l}={\l\over \bar\l}\\_{\l}-{1\over i\bar\l}\\_{\varphi}.$$
For an arbitrary holomorphic function $f(\l),\l\in \hbox{\Bbb C}
\backslash 0 $
 we have
$$\\_{\l}^n f(\l)=\bigl({1\over i\l}\\_{\varphi}\bigr)^nf(\l),\
\\_{\bar\l}^n\overline{f(\l)}=\bigl(-{1\over i\bar\l}\\_\varphi\bigr)^n
\overline{f(\l)},\ \l\in \hbox{\Bbb C}\backslash 0 \eqno (3.6)$$
So  we can replace the operators $\\_{\l}^m$ and $\\_{\bar\l}^m$ in (3.3)
by $\bigl({1\over i\l}\\_{\varphi}\bigr)^m$ and $\bigl(-{1\over i\bar\l}
\\_{\varphi}\bigr)^m$ respectively.

Comparing (1.18) and (3.3) we see that
$$\eqalignno{
a_m(\l(1+0))\bigg|_{|\l|=1}&=\bigl({1\over i\lp}\\_{\varphi^{\prime}}\bigr)^m
h_{-}(\l,\lp)\bigg|_{\lp=\l} &(3.7a-)\cr
b_m(\l(1+0))\bigg|_{|\l|=1}&=\bigl(-i\lp \\_{\varphi^{\prime}}\bigr)^m
h_{-}(\l,\lp)\bigg|_{\lp=-\l} &(3.7b-)\cr
a_m(\l(1-0))\bigg|_{|\l|=1}&=\bigl({1\over i\lp}\\_{\varphi^{\prime}}\bigr)^m
h_{+}(\l,\lp)\bigg|_{\lp=\l} &(3.7a+)\cr
b_m(\l(1-0))\bigg|_{|\l|=1}&=\bigl(-i\lp \\_{\varphi^{\prime}}\bigr)^m
h_{+}(\l,\lp)\bigg|_{\lp=-\l}. &(3.7b+)\cr  }  $$
 If (0.2),(0.5),(3.2) are fulfilled then the functions $f(\l,\lp),h_{\pm}
(\l,\lp)$ are M times continuously differentiable on the torus and
$b(\l)\in C_3^{(M)}(D_{-}).$

If $v(x)\in \prb$ then
$$f(\l,\lp)\in C^{(\infty)}(T^2), \eqno (3.8)$$
if, in addition, the "small norm" assumption (0.4) is valid then
$$b(\l)\in C_{\infty}^{(\infty)}(D_{\pm}). \eqno (3.9)$$
The definitions (3.3) and equations (3.5),(3.7), $m=0,\ldots,M$ and
the property (3.4),\ $m=0$ were given in [12].

Now we come up to one of the most important points of our paper. From
(3.4),(3.5),\par\noindent
(3.7) $m=0,\ldots,M$ we shall obtain $2M+2$ additional
necessary conditions on the scattering data $f(\l,\lp)$ and $b(\l)$
for a potential $v(x)$ with  properties (0.2),(0.5),(3.2).
A half of these conditions was given earlier in [9,\ 12]. Analogs of the
second half of these conditions for the case of negative energy were
 considered earlier in [11]. These conditions shall be written in terms
 of the functions $h_{-}(\l,\lp),b(\l)$.

We recall that the functions $h_-(\l,\lp)$ and $f(\l,\lp)$ are connected by
(1.28a).

Let us introduce new functions
$$a_m^{\pm}(\l)=\theta(\pm(1-\l\bar\l)) a_m(\l),\ b_m^{\pm}(\l)=\theta
(\pm(1-\l\bar\l))b_m(\l).\eqno (3.10)$$
$2M+2$ additional conditions on the scattering data will be obtained by
induction.

Let $M=0$. The equation (3.7a-) with $m=0$ takes the form
$$b_0^{-}(\l(1+0))\big|_{|\l|=1}=h_-(\l,-\l),\ b_0^-(\l)=\theta(\l\bar\l-1)
b_0(\l). \eqno (3.11)$$
The relation (3.11) is the first additional condition on the scattering data
$h_-(\l,\lp),b(\l).$
Let us calculate the function $a_0^-(\l)$ as a solution of the boundary
value problem for the equation (3.5a) in $D_-$ with the boundary conditions
(3.7a-) on the unit circle $|\l|=1$ and (3.4a-) on $\l=\infty$. This
boundary problem is solvable if and only if the following equality is valid
$$\displaystyle \bigl[{1\over {2\pi i}}\oint_{\\D_{-}}h_{-}(\xi,\xi)
{d\xi\over {\xi-\l}}-{1\over \pi}\intt_{D_-} {{\pi b(\xi)\overline{ b(\xi)}}
\over
\bar\xi}{{d\xi_Rd\xi_I}\over {\xi-\l}}\bigr]\bigg|_{|\l|=1-0}\equiv -s_0
\eqno (3.12)$$
for an appropriate constant $s_0$.
Under condition (3.12) the function $a_0^-(\l)$ takes the form
$$\displaystyle a_0^-(\l)={1\over {2\pi i}}\oint_{\\D_{-}}h_{-}(\xi,\xi)
{d\xi\over {\xi-\l}}-{1\over \pi}\intt_{D_-} {{\pi b(\xi)\overline{ b(\xi)}}
\over \bar\xi}
{{d\xi_Rd\xi_I}\over {\xi-\l}}+s_0.\eqno (3.13)$$
The relation (3.12) is the  second additional condition on the scattering
data
$h_-(\l,\lp),b(\l)$.

The step of induction is the following.

Let for a fixed $M=n$ we have found $2n+2$  additional conditions on the
scattering data and we have expressed the functions $a_m^-(\l),b_m^-(\l),
m=0,\ldots,n$ via $b_0^-(\l)$ and $\bigl({1\over i\lp}\\_{\varphi^{\prime}}
\bigr)^mh_-(\l,\lp)\big|_{\lp=\l},\ m=0,\ldots,n$.

Assume now that $M=n+1$.Then using equations (3.4-),(3.5),(3.7-),
 $m=0,\ldots,
n+1$ and expressions for $a_n^-(\l),b_n^-(\l)$ obtained at the previous step
we shall find 2 conditions more on the scattering data and we shall express
$a_{n+1}^-(\l),b_{n+1}^-(\l)$ via $b_0^-(\l)$,

\noindent
$\bigl({1\over i\lp}\\_{\varphi^{\prime}}
\bigr)^mh_-(\l,\lp)\big|_{\lp=\l},\ m=0,\ldots,n+1\  (C_{n+1}^{(3)}\prc
\subset C_n^{(3)}\prc $ so all conditions found for $M=n$ are fulfilled for
$M=n+1$).
Using (3.5b) we obtain $b_{n+1}^-(\l)$ as
$$b_{n+1}^-(\l)={\\ \over {\\ \bar\l}}b_n^-(\l)-{\pi\over \l}b_0^-(\l)
\overline{a_n^-(\l)}. \eqno (3.14)$$
The relation (3.7b-),\ $m=n+1$  is the $(2n+2+1)^{st}$ additional condition
on the scattering
 data
$$b_{n+1}^-(\l(1+0))\big|_{|\l|=1}=(-i\lp\\_{\varphi^{\prime}})^{n+1}
h_-(\l,\lp)\big|_{\lp=-\l}. \eqno (3.15)$$
The function $b_0^-(\l)\in C_3^{(n+1)}(\C)$ so the condition (3.4b-),
$m=n+1$ is fulfilled.

Let us calculate the function $a_{n+1}^-(\l)$ as a solution of the boundary
value problem for the equation (3.5a) in $D_-$ with    the boundary condition
(3.7a-) on the unit circle $|\l|=1$ and (3.4a-) on $\l=\infty$. This boundary
value problem is solvable if and only if the following equality is valid:
$$\displaystyle \eqalign{
\qquad \biggl[{1\over {2\pi i}}\oint_{\\D_{-}}&\biggl(\bigl(
{1\over i\lp}\\_{\varphi^{\prime}}\bigr)^{n+1}h(\xi,\lp)\big|_{\lp=\xi}
\biggr)
{d\xi\over {\xi-\l}}\cr
& -{1\over \pi}\intt_{D_-} {{\pi b_0^-(\xi)
\overline{b_{n+1}^-(\xi)}}\over \bar\xi}
{{d\xi_Rd\xi_I}\over {\xi-\l}}\biggr]\bigg|_{|\l|=1-0}\equiv -s_{n+1}
\cr} \eqno (3.16)$$
for an appropriate constant $s_{n+1}$.

Under condition (3.16) the function $a_{n+1}^-(\l)$ takes the form
$$\displaystyle  \eqalign{
  a_{n+1}^-(\l)=&{1\over {2\pi i}}\oint_{\\D_{-}}
\biggl(\bigl({1\over i\lp}\\_{\varphi^{\prime}}\bigr)^{n+1}h(\xi,\lp)
\big|_{\lp=\xi}
\biggr)
{d\xi\over {\xi-\l}}\cr
&-{1\over \pi}\intt_{D_-} {{\pi b_0^-(\xi)
\overline{b_{n+1}^-(\xi)}}\over \bar\xi}
{{d\xi_Rd\xi_I}\over {\xi-\l}}+s_{n+1}.\cr}\eqno (3.17)$$
The relation (3.16) is the $(2n+2+2)^{nd}$ additional condition on the
scattering data. We recall that the function $b_n^-(\l)$ is expressed via
$b_0^-(\l)$ , $\bigl({1\over i\lp}\\_{\varphi^{\prime}}
\bigr)^mh_-(\l,\lp)\big|_{\lp=\l},\ m=0,\ldots,n$. The step of induction is
done.

Thus an algorithm to write $2M+2$ additional conditions on the scattering
data for a potential with the properties (0.2),(0.5),(3.2) is presented.

In the paper [12] from equations (3.5),(3.7),$m=0,\ldots,M$ only ((3.4)
was not used) $M+1$ additional necessary conditions on the scattering
data were derived. These conditions can be considered as a method to
determine
$$\\_{\bar\l}^mb(\l)\bigl|_{|\l|=1+0},\ m=0,\ldots,M \eqno (3.18)$$
via the function $h_-(\l,\lp)$. The first of these conditions coincides
with (3.11). In [12] an algorithm  to write all these conditions was
suggested.
\medskip
\noindent
{\it Remark.}
Let $b(\l)$ be an arbitrary function such that $b(\l)\in C^{(M)}(D_-)$.
Then the derivatives (3.18) completely define all the derivatives
$$\\_{\bar\l}^{n_1}\\_{\l}^{n_2}b(\l)\big|_{|\l|=1+0} \eqno (3.19)$$
for all nonnegative integers $n_1,n_2$ such that $n_1+n_2\le M$.

These $M+1$ conditions on the scattering data are local for $b(\l)$
and almost local for $h_-(\l,\lp)$.We shall call these conditions local.
It is rather natural to replace $M+1$ conditions (3.11),(3.15) in the
family (3.11)-(3.17) by  local conditions. It can be shown that this new
collection of conditions is equivalent to the old one. The conditions
(3.12)-(3.14),(3.16),(3.17) are nonlocal for $b_0^-(\l),h_-(\l,\lp)$.

Thus for a potential with properties (0.2),(0.5),(3.2) $M+1$ additional
local conditions and $M+1$ additional nonlocal conditions on the
scattering data $b_0^-(\l),h_-(\l,\lp)$ are constructed.
\medskip
\noindent
{\it Remark.}
Analogs of these nonlocal conditions on the scattering data for a negative
energy were constructed earlier in [11].

We have studied boundary value problems on $D_-$. It is rather natural to
 consider analogs of these conditions on $D_+$. Let us show that these
new conditions are equivalent to the old ones.

According to (1.30) we have
$$h_+(\l,\lp)=\overline{h_-(-\lp,-\l)}. \eqno (3.20)$$
{\bf Lemma 1.}
{\it Let}\  $a_m(\l),b_m(\l)$ {\it be defined by the formulas } (3.3),
{\it where}

\noindent
$m=0,\ldots,M,\
v(x)$ {\it satisfy } (0.2),(3.2). {\it Then}
$$\displaystyle a_m^+(\l)=\sum_{k=0}^m\overline{\beta_{mk}(-1/\bar\l)}\
\overline{a_m^-(-1/\bar\l)}, \eqno (3.21a)$$
$$\displaystyle b_m^+(\l)=\sum_{k=0}^m\beta_{mk}(-1/\bar\l)\
\overline{b_m^-(-1/\bar\l)},
 \eqno (3.21b)$$
{\it where } $\beta_{mk}(\l)$ {\it are defined by}
$$\displaystyle (\l^2\\_{\l})^m=\sum_{k=0}^m\beta_{mk}(\l)\\_{\l}^k.
\eqno (3.22)$$
{\it The functions } $\beta_{mk}(\l)$ {\it have the following properties}
$$\eqalignno{
&{\rm a)}\ \beta_{mm}(\l)=\l^{2m}, &(3.23) \cr
&{\rm b)}\ \beta_{m0}(\l)=0\ {\it for}\ m>0, &(3.24) \cr
&{\rm c)}\ {\rm deg}\ \beta_{mk}(\l)=m+k\ {\it for}\ 0<k\le m, &(3.25) \cr
&{\rm d)}\ \beta_{m+1,k}(\l)=\l^2\beta_{m,k-1}(\l)+\l^2\\_{\l}b_{mk}(\l),
&(3.26)
\cr}$$
{\it where } $\beta_{00}=1,\ \beta_{m,-1}=0,\ \beta_{m,m+1}=0. $

The proof of lemma 1 follows from (3.3),(1.26) and the following relations.

Let
$$\eqalign{
F_m(\l)=\\_{\l}^m\exp\big[-{i\over 2}(\l\bar z+z/\l)\big], \cr
G_m(\l)=\\_{\bar\l}^m\exp\big[{i\over 2}(\bar\l z+\bar z/\bar\l)\big].\cr}
\eqno (3.27)$$
Then
$$\displaystyle \eqalign{
F_m(-1/\bar\l)=\sum_{k=0}^m\overline{\beta_{mk}(\l)}\,
\overline{F_k(\l)}, \cr
G_m(-1/\bar\l)=\sum_{k=0}^m\beta_{mk}(\l)
\overline{G_k(\l)}.\cr} \eqno (3.28)$$
\medskip
\noindent
{\bf Lemma 2.}
{\it Let the functions } $a_m^-(\l),b_m^-(\l),\ m=0,\ldots,M$ {\it satisfy
the boundary
value problem } (3.5),(3.4-),(3.7-). {\it Let the functions } $h_+(\l,\lp)$
{\it and}
$h_-(\l,\lp)$ {\it be connected by } (3.20). {\it Then the  functions }
$a_m^+(\l),
b_m^+(\l)$ {\it defined by } (3.21) {\it satisfy the boundary value problem }
(3.5),(3.4+),(3.7+).

The proof of this lemma will be given at the end of this section. Assume now
that $f(\l,\lp)\equiv 0$ (and $h_-(\l,\lp)\equiv 0$ accordingly).
Then the first $M+1$ local conditions simply mean that on the unit circle
$|\l|=1$ the function $b(\l)$ and all the derivatives $\\_{\l}^{n_1}\\_{
\bar\l}^{n_2}b(\l),\ n_1\ge 0,n_2\ge 0,\ n_1+n_2\le M$ are equal to zero.But
the nonlocal conditions in this case are rather nontrivial. The first  of
them
(3.12) takes the form
$$I_0(\l)\big|_{|\l|=1}\equiv -s_0, \eqno (3.29)$$
where
$$\displaystyle I_0(\l)=-{1\over \pi}\intt_{|\xi|\ge 1}{{\pi b(\xi)
\overline{ b(\xi)}}
\over \bar\xi}{{d\xi_Rd\xi_I}\over {\xi-\l}} \eqno (3.30)$$
and (3.13) takes the  form
$$a_0^-(\l)=I_0(\l)+s_0.  \eqno (3.31)$$
{}From (3.3a) it follows that
$$ a_0^-(\infty)=\hat v(0),  \eqno (3.32)$$
where $\hat v(p)$ is the Fourier transform of the potential $v(z)$.

{}From (3.30) it follows a rather interesting corollary.
\medskip
\noindent
{\bf Corollary 1.}
{\it Let } $v(z)$ {\it be a nonzero transparent } ({\it i.e. }
$f(\l,\lp)\equiv 0$) {\it at a fixed
energy } $E=1$ {\it potential satisfying } (0.2), {\it the "small norm"
condition } (0.5) {\it and}
$v(z)\in C_{2+\varepsilon}^{(3)}\prc$. {\it Then}
$$\hat v(0)>0,  \eqno (3.33)$$
{\it where}
$$\displaystyle \hat v(p)={\p}\intt_{z\in \C}e^{-{i\over 2}
(p\bar z +\bar p z)}
v(z)dz_Rdz_I,\ p\in \C. \eqno (3.34)$$

The proof of the corollary 1.

{}From (3.5a) it follows that
$$\\_{\bar\l}\,a_0^-(e^{i\varphi}\l)={{\pi \theta(\l\bar\l-1)}\over \bar\l}\,
|b(e^{i\varphi}\l)|^2 . \eqno (3.35)$$
Consider the average of $a_0^-(\l)$ over the angle
$$\displaystyle \alpha_0(\l)={1\over 2\pi}\int_0^{2\pi}a_0^-(e^{i\varphi}\l)
d\varphi. \eqno (3.36)$$
It has the following properties
$$\displaystyle \\_{\bar\l}\alpha_0(\l)={{\pi \theta(\l\bar\l-1)}\over
\bar\l}
{1\over 2\pi}\int_0^{2\pi}|b(e^{i\varphi}\l)|^2d\varphi, \eqno(3.37)$$
$\alpha_0(\l)=0\ {\rm as}\ |\l|\le 1,\ \alpha_0(\infty)=\hat v(0)$.

Consider the restriction of $\alpha_0(\l)$ on the real axis ${\rm Im}\
\l=0,\ {\rm Re}\ \l=r>0$. Then
$$\displaystyle \eqalign{
\\_r\alpha_0(r)=&2\\_{\bar\l}\alpha_0(\l)\big|_{{\rm Im}\ \l
=0} \cr
&={{2\pi}\over r}\theta(r^2-1)\int_0^{2\pi}|b(e^{i\varphi}r)|^2d\varphi
\ge 0, \cr} \eqno (3.38)$$
$\alpha_0(1)=0,\ \alpha_0(\infty)=\hat v(0)$. Thus,\ $\hat v(0)>0$.

{}From formula (3.32) and corollary 1 the corollary 2 follows.
\medskip
\noindent
{\bf Corollary 2.}
{\it Let the assumptions of corollary } 1 {\it be fulfilled. Then there
exists no
path connecting the points } 0 {\it and } $\infty$ {\it which has no
intersections
with the support of } $b(\l)$.

The proof of the corollary 2.

The function $a_0^-(\l)$ is identically equal to 0 as $|\l|\le 1$. If
such path exists then $a_0^-(\l)$ is holomorphic in a neighborhood of
this path and as a consequence identically equal to 0 along this path so
$a_0^-(\infty)=0$.  It contradicts to corollary 1.

Consider an  important particular class of potentials depending only on
$|z|,v(z)=v(|z|)$. In this case the functions $a_m(\l),b_m(\l)$ possess
the following symmetries.
\medskip
\noindent
{\bf Lemma 3.}
{\it Let the potential } $v(z)$ {\it depend only on } $|z|$, {\it i.e. }
$v(z)=v(|z|)$. {\it Then}
$$a_m(e^{i\varphi}\l)=e^{-im\varphi}a_m(\l),\ b_m(e^{i\varphi}\l)=
e^{im\varphi}b_m(\l). \eqno (3.39)$$
In this case all nonlocal conditions are fulfilled automatically.
\medskip
\noindent
{\bf Proposition 1.}
{\it Let the scattering data } $b(\l)$ {\it in the transparent case }
$f(\l,\lp)\equiv 0$
 {\it have the following properties}
$$\eqalign{
&1)\ b(\l)\in C_{\infty}^{(\infty)}(\C), \cr
&2)\ \\_{\l}^m\\_{\bar\l}^nb(\l)\big|_{|\l|=1}=0\ {\it for\  all}\ m,n\ge 0,
\cr
&3)\ b(e^{i\varphi}\l)=b(\l), \cr
&4)\ b(\l)=\overline{ b(\l)},\ b(1/\l)=b(\l)\ \cr
&\ \ \  ({\it it\  follows\  from\  property\ } 3\ {\it  and}\  (1.25)).
\cr}$$
{\it Then  all local and nonlocal conditions on the scattering data,
formulated
above, are fulfilled automatically}.

We shall prove proposition 1 by induction. The function $b_0^-(\l)=\theta(
\l\bar\l-1)b(\l)$ is known and satisfies the first additional condition.
The
step of induction is the following.

Suppose that under our assumption the first $2n+1$  additional conditions
on the scattering data are fulfilled, the functions $b_n^-(\l)$ and for
$n\ge 1\ \ a_{n-1}^-(\l)$ are expressed in terms
 of $b(\l)$ and these functions satisfy (3.39) and all derivatives  of
 $b_n^-(\l)$ vanish as $|\l|=1$.

We will show that two next  additional conditions are fulfilled, we will
express $b_{n+1}^-(\l)$ and $a_n(\l)$ in terms of $b(\l)$ and we will
see that these functions satisfy (3.39) and all derivatives of
$b_{n+1}^-(\l)$ vanish as $|\l|=1$.
We define
$$\displaystyle \eqalign{
a_n^-(\l)&={1\over \l^n}[I_n(\l)-I_n(1)], \cr
I_n&=\\_{\bar\l}^{-1}\bigg[{{\pi \l^n}\over \bar\l}\theta(\l\bar\l-1)
b(\l)\overline{ b_n(\l)}\bigg]= \cr
&-{1\over \pi}\intt_{\C}{{\pi \xi^n}\over \bar\xi}\theta(\xi\bar\xi-1)
{{b(\xi)\overline{ b_n(\xi)}}\over {\xi-\l}}d\xi_Rd\xi_I. \cr} \eqno (3.40)$$
The function $I_n(\l)$ is well-defined and $I_n(\l)=I_n(|\l|)$ (it follows
from (3.39)). Thus, $a_n^-(\l)\equiv 0$ as $|\l|=1$ and it solves the
boundary
value problem (3.5a),(3.7a-) with $h_-(\l,\lp)\equiv 0$,(3.4a-) and it
 satisfies (3.39). Now we can define
$$b_{n+1}^-(\l)=\\_{\bar\l}b_n^-(\l)-{\pi\over \bar\l}b(\l)
\overline{ a_n^-(\l)}.
\eqno (3.41)$$
We see that if $b_n^-(\l)$ satisfies (3.7b-) and all derivatives of
$b_n^-(\l)$
vanishes as $|\l|=1$ then the same is valid for $b_{n+1}^-(\l)$. It is
the step
of induction.

The proof is completed.

{\it The proof of lemma } 2.

{}From (3.21) and (3.5) in $D_-$ and (3.23),(3.26) we obtain the following
 relations in $D_+\backslash 0 $
$$\eqalignno{
&\\_{\bar\l}a_m^+(\l)=\\_{\bar\l}\sum_{k=0}^m\overline{\beta_{mk}(-1/\bar\l)}
\ \overline{a_k^-(-1/\bar\l)}
=\sum_{k=0}^m\overline{\beta_{mk}(-1/\bar\l)}
\ \overline{\\_{\l}a_k^-(-1/\bar\l)} \cr
&=\sum_{k=0}^m\overline{\beta_{mk}(-1/\bar\l)}
\ \overline{[\bar\mu^2\\_{\bar\mu}a_k^-(\mu)]}\big|_{\mu=-1/\bar\l} \cr
&=\sum_{k=0}^m\overline{\beta_{mk}(-1/\bar\l)}\ {1\over \bar\l^2}
\pi\ {\rm sgn}\,\bigl({1\over \l\bar\l}-1\bigr) (-\bar\l)
\ \overline{b(-1/\bar\l)}\ b_k^-(-1/\bar\l) \cr
&={{\pi\ {\rm sgn}\,(\l\bar\l-1)}\over \bar\l}b(\l)\overline{
\sum_{k=0}^m\beta_{mk}(-1/\bar\l)\ \overline{b_{k}^-(-1/\bar\l)}}=r(\l)
\ \overline{b_m^+(\l)}, &(3.42) \cr} $$
$$\eqalignno{
&\\_{\bar\l}b_m^+(\l)=\\_{\bar\l}\sum_{k=0}^m\beta_{mk}(-1/\bar\l)
\ \overline{b_k^-(-1/\bar\l)}
=\sum_{k=0}^m\beta_{m+1,k}(-1/\bar\l)
\ \overline{b_k^-(-1/\bar\l)} \cr
&-{1\over \bar\l^2}\sum_{k=0}^m\beta_{m,k-1}(-1/\bar\l)
\ \overline{b_k^-(-1/\bar\l)}+\sum_{k=0}^m\beta_{mk}(-1/\bar\l)
\ \overline{(\\_{\l}b_k^-(-1/\bar\l))} \cr
&=\sum_{k=0}^m\beta_{m+1,k}(-1/\bar\l)
\ \overline{b_k^-(-1/\bar\l)}-{1\over \bar\l^2}
\ \sum_{k=0}^m\beta_{m,k-1}(-1/\bar\l)
\ \overline{b_k^-(-1/\bar\l)} \cr
&+ \sum_{k=0}^m\beta_{mk}(-1/\bar\l)
\ \overline{[\bar\mu^2\\_{\bar\mu}b_k^-(\mu)]}\big|_{\mu=-1/\bar\l}=
\ \sum_{k=0}^m\beta_{m+1,k}(-1/\bar\l)
\ \overline{b_k^-(-1/\bar\l)} \cr
&-{1\over \bar\l^2}\sum_{k=0}^m\beta_{m,k-1}(-1/\bar\l)
\ \overline{b_k^-(-1/\bar\l)}+
{1\over \bar\l^2}\sum_{k=0}^m\beta_{mk}(-1/\bar\l)
\ \overline{b_{k+1}^-(-1/\bar\l)} \cr
&+\sum_{k=0}^mb_{mk}(-1/\bar\l)\ {1\over \bar\l^2}\,\pi\ {\rm sgn}
\,\bigl({1\over \l\bar\l}-1\bigr)\ (-\bar\l)\ \overline{ b(-1/\bar\l)}
\ a_k^-(-1/\bar\l) \cr
&=\sum_{k=0}^m\beta_{m+1,k}(-1/\bar\l)
\ \overline{b_k^-(-1/\bar\l)}+
{{\beta_{mm}(-1/\bar\l)}\over \bar\l^2}
\ \overline{b_{m+1}^-(-1/\bar\l)}
+r(\l)\ \overline{a_m^+(\l)} \cr
&=b_{m+1}^+(\l)+r(\l)
\ \overline{a_m^+(\l)}. &(3.43) \cr} $$
Thus, the functions $a_m^+(\l),b_m^+(\l)$ satisfy (3.5) in $D_+\backslash 0$.
The relations (3.4+) follow from (3.4-) and (3.23),(3.24),(3.25).

To prove (3.7+) we use the following identities
$$\eqalign{
\bigl({1\over i\lp}\\_{\varphi^{\prime}}\bigr)^m&=
\sum_{k=0}^m\overline{\beta_{mk}(-1/\bar\l^{\prime})}
\ (-i\lp\\_{\varphi^{\prime}})^k, \cr
(-i\lp\\_{\varphi^{\prime}})^m&=
\sum_{k=0}^m\beta_{mk}(\lp)
\ \bigl({1\over i\lp}\\_{\varphi^{\prime}}\bigr)^k. \cr} \eqno (3.44)$$
The relations (3.44) follow from (3.23),(3.26).

Due to (3.21a),(3.7a-),(3.20),(3.44) we have
$$\eqalign{
a_m^+(\l(1-0))\big|_{|\l|=1}&=
\sum_{k=0}^m\overline{\beta_{mk}(-1/\bar\l)}
\ \overline{\bigl({1\over i\lp}\\_{\varphi^{\prime}}\bigr)^k
h_-(-\l,-\lp)}\big|_{\lp=\l} \cr
&=\sum_{k=0}^m\overline{\beta_{mk}(-1/\bar\l^{\prime})}
\ (-i\lp\\_{\varphi^{\prime}})^k
\ h_+(\l,\lp)\big|_{\lp=\l} \cr
&=\bigl({1\over i\lp}\\_{\varphi^{\prime}}\bigr)^m
\ h_+(\l,\lp)\big|_{\l=\lp}\ , \cr
b_m^+(\l(1-0))\big|_{|\l|=1}&=
\sum_{k=0}^m\beta_{mk}(-1/\bar\l)
\ \overline{(+i\lp\\_{\varphi^{\prime}})^k
\ h_-(-\l,-\lp)}\big|_{\lp=-\l}  \cr
&=\sum_{k=0}^m\beta_{mk}(\lp)
\ \bigl({1\over i\lp}\\_{\varphi^{\prime}}\bigr)^k
\ h_+(\l,\lp)\bigl|_{\lp=\l} \cr
&=(-i\lp\\_{\varphi^{\prime}})^m
\ h_+(\l,\lp)\big|_{\lp=\l}\ .\cr} \eqno (3.45)$$
Lemma 2 is proved.
\vfill\eject
\noindent
{\bf 4.The construction of potentials with zero scattering amplitude at fixed
 energy.}
\medskip
\noindent
Now we are ready to formulate one of the main results of our paper.
\medskip
\noindent
{\bf Theorem 1.}
{\it Let } $b(\l),\ \l\in \C$ {\it be an arbitrary function  with the
following properties}:

l)\ $b(\l)\in C_{\infty}^{(\infty)}(\C)$. \par
2)\ $b(1/\bar\l)=b(\l),\ b(-1/\bar\l)=\bar b(\l)$. \par
3)\ $\\_{\l}^m \\_{\bar\l}^n b(\l)\big|_{|\l|=1}\ =0\ {\it for\ all}
\ m,n\ge 0.  $ \par
4)\ {\it The function } $b(\l)$ {\it satisfies the first } $M+1$
{\it nonlocal conditions on the
scattering data formulated in the previous section for the case } $f(\l,\lp)
\equiv 0$,\ {\it i.e. the following } $M+1$ {\it boundary value problems on }
$D_-=\{ \l\in
\C\ |\ |\l|\ge 1 \} $
$$\eqalign{
&\\_{\bar\l}a_m^-(\l)=r(\l)\,\overline{b_m^-(\l)}, \cr
&a_m^-(\l)\big|_{|\l|=1}=0,\ a_m^-(\infty)=O(1),\ m=0,1,\ldots,M \cr}
\eqno (4.1)$$
{\it are resolvable, where the functions } $b_m^-(\l)$ {\it are defined
recurrently by}:
$$\eqalignno{
b_{m+1}^-(\l)&=\\_{\bar\l}b_m^-(\l)-{\pi\over \bar\l}\theta(\l\bar\l-1)\,
b(\l)
\,\overline{ a_m^-(\l)}, &(4.2)\cr
b_0^-(\l)&=\theta(\l\bar\l-1)\,b(\l). &(4.3)\cr}$$
{\it Then the potential } $v(z)$ {\it constructed from the  scattering data }
$b(\l),
f(\l,\lp)\equiv 0$ {\it by the procedure, described in section } 2 {\it for }
$E=1$ {\it has the
following properties}:

1)$v(z)$ {\it is real-valued}. \par
2)$v(z)\in C_{M+3}^{(\infty)}\prc$. \par
3){\it The scattering amplitude for the two-dimensional Schr\"odinger
equation}
$$-4\\_z\\_{\bar z}\psi(z,\l)+v(z)\psi(z,\l)=E\psi(z,\l),\ E=1 \eqno (4.4)$$
{\it is equal to zero } $(f(\l,\lp)\equiv 0\ |\l|=|\lp|=1)$ {\it at the
energy level}
$E=1$. {\it Moreover, the classical scattering solutions } $\varphi^+(z,\l)$
{\it of}
(4.4) {\it have the following asymptotics}
$$\varphi^+(z,\l)=e^{{i\over 2}(\l\bar z +z/\l)}+O\bigl(|z|^{-M-2}
\bigr),\ {\it where }\ |\l|=1, \eqno (4.5)$$
{\it or in the standard notation}
$$\varphi^+(k,x)=e^{ikx}+O\bigl(|x|^{-M-2}\bigr),\ k^2=E=1. \eqno
(4.6) $$
{}From theorem 1 and proposition 1 it follows:
\medskip
\noindent
{\bf Theorem 2.}
{\it Let } $f(\l,\lp)\equiv 0,b(\l)$ {\it satisfy the same conditions as
in proposition }1.
{\it Then the corresponding real potential } $v(z)\in C_{\infty}^{(\infty)}
\prc$
{\it and}
$$\varphi^+(k,x)=e^{ikx}+O\bigl(1/|x|^{\infty}\bigr),\ k^2=E=1$$
({\it i.e. } $\varphi^+(k,x)-e^{ikx}$ {\it decays as } $|x|\to \infty$
{\it faster then
any degree of } $|x|^{-1}$).

The proof of  theorem 2.
If the scattering amplitude $f(\l,\lp)\equiv 0$ (and $\rho(\l,\lp)\equiv 0$
accordingly) then the function $\mu(z,\l)$ (see section 2) is defined as
a solution of the equation
$$\\_{\bar\l}\mu(z,\l)=r(\l,z)\overline{\mu(z,\l)}, \eqno (4.7)$$
such that
$$\mu(z,\l)\to 1\ {\rm as}\ |\l|\to \infty, \eqno (4.8)$$
where
$$r(\l,z)=e^{-{i\over 2}(\l\bar z +z/\l+\bar\l z+\bar z/\bar\l)}r(\l),\
\ r(\l)={{\pi\ {\rm sgn}\,(\l\bar\l-1)}\over \bar\l}b(\l), \eqno (4.9)$$
or equivalently\ the function $\mu(z,\l)$ is defined as a solution of
the integral equation
$$\mu(z,\l)=1+(A_z\mu)(z,\l), \eqno (4.10) $$
where
$$\displaystyle (A_zf)(\l)=\\_{\bar\l}^{-1}(r(\l,z)\overline{ f(\l)})=-
{1\over \pi}
\intt_{\C}{{r(\zeta,z)\overline{ f(\zeta)}}\over {\zeta-\l}}d\zeta_Rd\zeta_I,
\eqno (4.11)$$
or equivalently
$$\mu(z,\l)=1+A_z\cdot 1+(A_z^2\mu)(z,\l). \eqno (4.12)$$
According to the theory of generalized analytic functions (see [34]) the
equations  (4.10),
\par\noindent
(4.12) have unique solution for all $z$.

This solution can be written as
$$\mu(z,\l)=\bigl(I-A_z^2\bigr)^{-1}(1+A_z\cdot 1). \eqno (4.13)$$
The equation (4.13) possesses a   formal asymptotic expansion
$$\mu(z,\l)=\bigl(I+A_z^2+A_z^4+A_z^6+\ldots\bigr)(1+A_z\cdot 1). \eqno
(4.14)$$
{}From (4.22) it follows that (4.14) uniformly converges for sufficiently
large $|z|$.

To study (4.14) we need some estimates on $A_z^2,A_z$. It is convenient
to write $A_z^2$ as
$$\eqalignno{
(A_z^2f)(z,\l)&={1\over \pi^2}\intt_{\C} K(z,\l,\eta)f(\eta)
d\eta_Rd\eta_I,\ {\rm where}\ &(4.15) \cr
K(z,\l,\eta)&=I(\l,\eta,z)\exp\big[{i\over 2}(\eta \bar z+z/\eta+\bar\eta z
+\bar z/\bar\eta)\big]\overline{ r(\eta)}, &(4.16)\cr
I(\l,\eta,z)&=\intt_{\C}{{r(\zeta)}\over {(\zeta-\l)(\bar\eta-\bar\zeta)}}
\exp\big[-{i\over 2}(\zeta\bar z+z/\zeta+\bar\zeta z+\bar z/\bar\zeta)\big]
d\zeta_Rd\zeta_I. &(4.17) \cr}$$
\medskip
\noindent
{\bf Lemma 4.}
{\it Let } $b(\l)$ {\it satisfy the conditions } 1),2),3)  {\it  of
theorem } 1. {\it Then}
\item{1)} $$\big|\\_z^{m}\\_{\bar z}^{n} I(z,\l,\eta)\big|\le
{\alpha_{mn}^{(1)}\over (1+|z|)}\,{{(1+\varphi(|\l-\eta|))}\over |\l-\eta|},
\eqno (4.18)$$
{\it where } $\varphi(z)={\ln r\over {1+r|\ln r|}}$ {\it for all } $
m,n\ge 0 $.

\item{2)} {\it For an arbitrary testing function } $f(\l)\in C(\bar{\C})$
{\it we have } $(A_zf)\in C(\bar{\C}),\ (A_z^2f)\in C(\bar{\C})$ {\it and
the following estimates
are valid}

\itemitem{a)} $$\eqalignno{
(A_z^2f)(\l)&={c_{-1}(z)\over \l}+O\bigl({1\over |\l|^2}\bigr)
\ {\rm for}\ \l\to\infty,\ {\rm where}\ &(4.19) \cr
c_{-1}&=-{1\over \pi^2}\intt_{\C}\intt_{\C}{{r(\zeta)\overline{r(\eta)}}\over
{\bar\eta-\bar\zeta}}\exp\big[-{i\over 2}(\zeta\bar z +z/\zeta +\bar\zeta
z+\bar z/\bar\zeta)\big] \cr
&\times\exp\big[{i\over 2}(\eta\bar z +z/\eta +\bar\eta z +
\bar z/\bar\eta)\big]
f(\eta)d\zeta_Rd\zeta_Id\eta_Rd\eta_I, &(4.20)\cr
\big|\\_z^m\\_{\bar z}^n\,c_{-1}\big|&\le {\beta_{mn}^{(1)}\over (1+|z|)}
\|f\|_C \ {\rm for\ all}\ m,n\ge 0.  &(4.21)\ \cr}  $$
\itemitem{b)} $$\big\|\bigl(\\_z^m\\_{\bar z}^nA_z^2\bigr)f\big\|_C\le
{\gamma_{mn}^{(1)}\over (1+|z|)}\|f\|_C\ {\rm for\ all}\ m,n\ge 0. \eqno
(4.22)$$
\itemitem{c)} $$\big\|\bigl(\\_z^m\\_{\bar z}^nA_z\bigr)f(\l)\big\|_C\le
\varepsilon_{mn}\,\|f\|_C\ {\rm for\ all}\ m,n>0. \eqno (4.23)$$
\par
(4.14) may be written as
$$\mu(z,\l)=1+A_z\cdot 1 +A_z^2\cdot 1 +\ldots +A_z^{2M+5}\cdot 1 +
R_M, \eqno (4.24)$$
where
$$\displaystyle R_M=\Biggl(\sum_{k=M+3}^{\infty}A_z^{2k}\Biggr)
(1+A_z\cdot1). \eqno (4.25)$$
{}From lemma 4 we get the following estimates on $R_M$.
\medskip
\noindent
{\bf Lemma 5.}
{\it Let } $b(\l)$ {\it satisfy conditions } 1),2),3) {\it of theorem } 1.
{\it Then }
\item{1)} $$\big|\\_z^m\\_{\bar z}^nR_M\big|\le
{\alpha_{mn}^{(2)}\over (1+|z|)^{M+3}}\  {\it for\ all}\ m,n\ge 0. \eqno
(4.26)$$
\item{2)} $$ R_M={q(z)\over \l}+O\bigl({1\over \l^2}
\bigr),\ {\it as} \l\to\infty, {\it where}\ \eqno (4.27)$$
$$\big|\\_z^m\\_{\bar z}^nq(z)\big|\le {\beta_{mn}^{(2)}\over (1+|z|)^{M+3}}\
\ {\it for\ all}\ m,n\ge 0. \eqno (4.28)$$
{}From (2.4),(2.5),(4.28) we see that the term $R_M$ gives a contribution to
the potential $v(z)$ and to the function $\mu(z,\l)$ from the functional
class $C_{M+3}^{(\infty)}\prc.$
\medskip
\noindent
{\bf Lemma 6.}
{\it To calculate the potential } $v(z)$ {\it and the function } $\mu(z,\l)$
{\it up to terms
of the order } $O\bigl(1/|z|^{M+3}\bigr)$ {\it it is sufficient to consider
only the
first } $2M+6$ {\it terms in the  formula } (4.24).

Let
$$\mu_M(z,\l)=1+A_z\cdot 1 +\ldots +A_z^{2M+5}\cdot 1. \eqno (4.29)$$
Now we give some estimates of $\mu_M(z,\l)$.
\medskip
\noindent
{\bf Lemma 7.}
{\it Let the conditions } 1)-3) {\it of theorem } 1 {\it be valid, }
$f(z,\l)$ {\it be a smooth function
of } $\l,z$ {\it such that all the derivatives } $\\_z^k\\_{\bar z}^l\\_
{\l}^m
\\_{\bar\l}^nf(z,\l)$ {\it are bounded on the } $\l$-{\it plane uniformly
in } $z$, {\it i.e.}
$$\big|\\_z^k\\_{\bar z}^l\\_{\l}^m\\_{\bar\l}^nf(z,\l)\big|\le
\alpha_{klmn}
\ {\it for\  all}\ z,\l. \eqno (4.30)$$
{\it Then}
\item{1)} $$\eqalign{
A_z*f(z,\l)&={2i\over (z-\bar z/\bar\l^2)}\exp\big[-{i\over 2}
(\l\bar z +\bar\l z+ z/\l +\bar z/\bar\l)\big] \cr
&\times\Biggl\{\Bigg[\sum_{k=0}^{\infty}(-1)^k\bigl(\\_{\bar\l}\circ
 {2i\over {z-\bar z/\bar\l^2}}\bigr)^k\Bigg]*(r(\l)\overline{ f(z,\l)})+
O\bigl({1\over |z|^{\infty}}
\bigr)\Biggr\}.\cr} \eqno (4.31)$$
{\it Here } $\circ$ {\it denotes the product of operators and } $*$
{\it means that we apply a
differential operator to the function}.
\item{2)} {\it Consider the asymptotic expansion of } $A_z*f(z,\l)$
{\it as } $\l\to\infty$
$$A_z*f(z,\l)={{\alpha_{-1}(z)}\over \l}+{{\alpha_{-2}(z)}\over \l^2}
+\ldots \eqno (4.32)$$
($r(\l)$ {\it vanishes as } $|\l|\to\infty$, {\it so we have no
nonholomorphic terms in}
(4.32)). {\it Then all } $\alpha_{-k}(z)\in C_{\infty}^{(\infty)}\prc$.
\item{3)} {\it Let } $|\l|=1$. {\it Then the function } $A_z*f(z,\l)$ {\it
decreases as}
$z\to\infty$ {\it faster than any degree of } $|z|$ {\it together with all
her
 derivatives}.
\item{4)} $A_z^2*f(z,\l)$ {\it is a smooth function of} $z,\l$
{\it such that}  $|z|\big|\bigl(\\_z^{n_1}\\_{\bar z}^{n_2}\\_{\l}^{k_1}
\\_{\bar\l}^{k_2}
f(z,\l)\bigr)\big|$ {\it are bounded in } $z,\l$. ({\it For } $A_z*f(z,\l)$
{\it it is not true}.)
\item{5)} $$\displaystyle
 A_z^2*f(z,\l)=-\\_{\bar\l}^{-1}\Biggl\{r(\l){1\over wR}
\Biggl[\sum_{k=0}^{\infty}{1\over w^k}\bigl(\\_{\l}\circ {1\over R}\bigr)^k
\Biggr]*(\overline{ r(\l)}f(\l,z))\Biggr\}+O\bigl({1\over
|z|^{\infty}}\bigr),
\eqno (4.33)$$
{\it where}
$$w=\bar z/2i,\ R=(1-\nu/\l^2),\ \nu=z/\bar z.
 \eqno (4.34)$$
\medskip
\noindent
{\bf Lemma 8.}
{\it Consider the function } $A_z^{2k+1}*1 $, {\it where} $k\in\N\cup 0$.
{\it Then}
\item{1)} {\it For } $|\l|=1$
$$ A_z^{2k+1}*1\in C_{\infty}^{(\infty)}\prc\ {\it in}\ z.$$
\item{2)} {\it Consider the asymptotic expansion of this function as }
$\l\to\infty$
$$A_z^{2k+1}*1={{\chi_{-1,2k+1}(z)}\over \l} +O\bigl({1\over \l^2}\bigr).$$
{\it Then } $\chi_{-1,2k+1}(z)\in C_{\infty}^{(\infty)}\prc\ {\it in}\ z.$
\par
So, all the terms $A_z^{2k+1}*1,\ k\in \N\cup 0$ in expansion (4.24) give a
contribution to the function $\mu(z,\l)$ for $|\l|= 1$ and to the potential
$v(z)$ defined by (2.5) from the functional class $C_{\infty}^{(\infty)}\prc$
in $z$ and in the asymptotic calculations these terms can be neglected.

This statement directly follows from statements 2-4 of lemma 7.

We have proved that if we want to calculate the potential $v(z)$ up to
terms of
the order $O\bigl(1/|z|^{M+3}\bigr)$ it is sufficient to approximate the
 function $\mu(z,\l)$ by
$$\mu(z,\l)\sim \mu_M^{\rm Appr}(z,\l)=1+A_z^2\cdot 1 +A_z^4\cdot 1 + \ldots
+
A_z^{2M+4}\cdot 1. \eqno (4.35)$$
To calculate the asymptotic expansion of $\mu_M^{\rm Appr}(z,\l)$ for large
$|z|$ let us apply the formula (4.33) from lemma 7.

The direct calculation with help of the formula (4.33) shows that
$$\mu_M^{\rm Appr}(z,\l)=1+{{{\cal C}_0(\l,\nu)}\over w}+
{{{\cal C}_1(\l,\nu)}
\over w^2}+\ldots
+{{{\cal C}_M(\l,\nu)}\over w^{M+1}}+{{{\cal C}_{M+1}(\l,\nu)}\over w^{M+2}}
+u_M(z,\l), \eqno (4.36)$$
where $w,R,\nu$ are defined by (4.34),\
$$u_M(z,\l)\in C_{M+3}^{(\infty)}(\R^2\backslash D)\ {\rm in}\  z\ {\rm  for
\ all}\ \l, \eqno (4.37)$$
$$u_M(z,\l)=u_{M,-1}(z)/\l+O(1/|\l|^2)\ {\rm as}\ \l\to\infty,
\ u_{M,-1}(z)\in C_{M+3}^{(\infty)}\in (\R^2\backslash D),\eqno (4.38)$$
$D$ is the unit disc $|z|<1$,\ the functions ${\cal C}_k(\l,\nu)$ are defined
 recurrently by
$$\displaystyle {\cal C}_n(\l,\nu)=-\\_{\bar\l}^{-1}\Biggl[{1\over R}r(\l)
\sum_{k=0}^n
\bigl(\\_{\l}\circ {1\over R}\bigr)^k*\bigl[\overline{ r(\l)}
{\cal C}_{n-k-1}(\l,\nu)
\bigr]\Biggl],\ {\cal C}_{-1}(\l,\nu)\equiv 1. \eqno (4.39)$$
Really, we have
$$\mu_{M+1}^{\rm Appr}(z,\l)=1+A_z^2*\mu_M^{\rm Appr}(z,\l). \eqno (4.40)$$
Substituting (4.33) to (4.40) and comparing expansion coefficients at
$1/w^{n+1},\ n=0,\ldots,M+1$ in both sides we get (4.39).

Let us introduce some additional notations. Consider the differential
operator

${1\over R}\bigl(\\_{\l}\circ {1\over R}\bigr)^n$. It can be written as
$$\displaystyle {1\over R}\bigl(\\_{\l}\circ {1\over R}\bigr)^n=\sum_{k=0}^n
f_{nk}\\_{\l}^k,\ {\rm where}\ f_{nk}=f_{nk}(\l,\nu),R=R(\l,\nu)=1-\nu/\l^2.
\eqno (4.41)$$
The functions $f_{nk}=f_{nk}(\l,\nu)$ have the following properties:
\itemitem{a)}
$$f_{nn}={1\over R^{n+1}}={1\over (1-\nu/\l^2)^{n+1}}. \eqno (4.42)$$
\itemitem{b)}
$$f_{nk}=\delta_{nk}+O\bigl(1/|\l|^{n-k+2}\bigr)\ {\rm as}\ \l\to\infty.
\eqno (4.43)$$
\itemitem{c)}
$$f_{nk}=O(|\l|^{n+k+2})\ {\rm as} \ \l\to 0. \eqno (4.44)$$
\itemitem{d)}
$$\eqalign{
f_{n+1,k}&={1\over R}(\\_{\l}f_{nk}+f_{n,k-1}),\ f_{00}={1\over R}=
{1\over {1-\nu/\l^2}},\cr
f_{k,-1}&=0,\ f_{k,k+1}=0. \cr}\eqno(4.45)$$
\itemitem{e)}For a fixed $\nu \ f(\l,\nu)$ is meromorphic in $\l$ with
poles only
 in points $\l^2=\nu$. In these points
$$f_{nk}=O\bigl({1\over {\l^2-\nu}}\bigr)^{2n-k+1}.\eqno (4.46)$$
\noindent\noindent
Now we can formulate the main algebraic lemma of our article.
\medskip
\noindent
{\bf Lemma 9.}
{\it Let the scattering data } $b(\l)$ {\it satisfy conditions of theorem }1.
{\it Then for}

\noindent
$n=1,\ldots,M$ {\it we have}
$$\displaystyle {\cal C}_n(\l,\nu)=\sum_{l=0}^nf_{nl}c_l(\l), \eqno(4.47)$$
{\it where functions } $c_l(\l)$ {\it satisfy the following equation}
$$\displaystyle \\_{\bar\l}c_l(\l)=-\Biggl[r(\l)\sum_{k=0}^n\\_{\l}^k
(\overline{ r(\l)}c_{n-k-1}(\l))\biggr],\ c_{-1}(\l)\equiv 1,\ \l\ne 0.
\eqno(4.48)$$
({\it It can be obtained by a formal substitution } $\nu=0$
{\it in equation } (4.39)). {\it These
functions do not depend on } $\nu$ {\it and are defined by}
$$\displaystyle c_n(\l)=-\pi {\rm sgn}\,(\l\bar\l-1)\sum_{k=0}^n\alpha_{nk}
(\l)a_k(\l),\eqno(4.49)$$
{\it where } $\alpha_{nk}(\l)$ {\it are defined by}
$$\displaystyle \\_{\l}^n\circ {1\over \l}=\sum_{k=0}^n\alpha_{nk}(\l)\circ
\\_{\l}^k,\ \alpha_{nk}(\l)=(-1)^{n-k}{{n!}\over {k!}}{1\over \l^{n-k+1}}\
\eqno(4.50)$$
{\it and functions } $a_m(\l)$ {\it have the form}
$$a_m(\l)=\theta(\l\bar\l-1)a_m^-(\l)+\theta(1-\l\bar\l)a_m^+(\l),
\eqno(4.51)$$
{\it where } $a_m^-(\l)$ {\it are defined by the boundary value problems }
(4.1) {\it and}
$a_m^+(\l)$ {\it are defined by } (3.21a).

The proofs of lemmas 4,5,7,9 will be given at the end of this section.

Using lemmas 4,5,8,9 we can complete the proof of theorem 1.

The fact that the reconstruction procedure from the scattering data
described
in section 2 gives real smooth nonsingular potentials was proved in the
 previous papers.

Using (4.36),(4.47),(4.49),(4.43),(4.50),(4.34) we obtain the following
estimate on $\mu_M^{\rm Appr}(z,\l)$ as $\l\to\infty$
$$\displaystyle \mu_M^{\rm Appr}(z,\l)=1-{\pi\over \l}\sum_{k=0}^M
\bigl({2i\over {\bar z}}\bigr)^{k+1}a_k(\infty)+{{(2i)^{M+2}{\cal C}_
{M+1,-1}(\nu)}
\over
{\bar z^{M+2}}}\,{1\over \l}+{{u_{M,-1}(z)}\over \l}+O\bigl({1\over |\l|^2}
\bigr).\eqno(4.52)$$
Here,
$${\cal C}_{M+1}(\l,\nu)={{{\cal C}_{M+1,-1}(\nu)}\over \l}+O\bigl({1\over
|\l|^2}\bigr).\eqno(4.53)$$
{}From (4.52),(2.5),(4.34) and lemmas 6,8 it follows that
$$v(z)={{(2i)^{M+3}}\over {\bar z^{M+3}}}\,{{\\ {\cal C}_{M+1,-1}(\nu)}\over
 {\\ \nu}}+2i\\_z u_{M,-1}(z)+\tilde v_M(z),\eqno(4.54)$$
where $\tilde v_M(z)\in C_{M+3}^{(\infty)}(\R^2\backslash D)$, all the terms
in (4.54) are from the functional class $C_{M+3}^{(\infty)}
(\R^2\backslash D)$.
The function $v(z)$ is smooth, so the statement 2 is proved.

{}From (4.1) and property 3) of the function $b(\l)$ in the formulation of
theorem 1 it follows that
$$\\_{\l}^{n_1}\\_{\bar\l}^{n_2}a_m^-(\l)\big|_{|\l|=1}=0\ {\rm for\ all}\
n_1,n_2\ge 0.\eqno(4.55)$$
Using (4.36),(4.47),(4.49), property e) of the function $f_{nk}$, (4.50),
(4.55) and lemma 8 we see that
$${\cal C}_k(\l,\nu)\big|_{|\l|=1}\equiv 0,\ k=0,\ldots,M \eqno(4.56)$$
and
$$\mu(z,\l)\big|_{|\l|=1}=1+{{{\cal C}_{M+1}(\l,\nu)}\over
w^{M+2}}+u_M(z,\l)+
\tilde\mu_M(z,\l),\eqno(4.57)$$
where

$u_M(z,\l)\in C_{M+3}^{(\infty)}(\R^2\backslash D),\ \tilde\mu_M(z,\l)\in
C_{M+3}^{(\infty)}(\R^2\backslash D).$

{}From (4.57) it follows that the function
$$\varphi(z,\l)=e^{{i\over 2}(\l\bar z+z/\l)}\mu(z,\l)\big|_{|\l|=1}
 \eqno(4.58)$$
has asymptotics (4.5). Now using rather standard arguments we show that
$\varphi (z,\l)$ coincides with the physical solution $\varphi^+(z,\l)$.
Theorem 2 is proved.
\medskip
{\it The scheme of the proof of Lemma } 4. We prove the estimate (4.18) only.
The estimate (4.22) follows directly from (4.18) and conditions 1),2),3)
of theorem 1. The estimates (4.19),(4.21),(4.22) are rather simple.

For the sake of definiteness let us assume that $m=n=0$.(The proof for
general $m,n\ge 0$ is very similar).

The denominator in (4.17) can be transformed in the following way
$$\eqalign{
{1\over (\zeta-\l)(\bar\eta-\bar\zeta)}&=-{1\over (\zeta-\l)(\zeta-\eta)}
{(\eta-\zeta)\over (\bar\eta-\bar\zeta)}={1\over {\eta-\l}}\bigl({1\over
{\zeta-\l}}-{1\over {\zeta-\eta}}\bigr)\ {(\eta-\zeta)\over (\bar\eta-
\bar\zeta)}\cr
&={1\over {\eta-\l}}\bigl({1\over {\bar\eta-\bar\zeta}}+{1\over {\zeta-\l}}
{{\eta-\zeta}\over {\bar\eta-\bar\zeta}}\bigr).\cr}\eqno(4.59)$$
According to (4.59) we have
$$I(\l,\eta,z)={1\over {\eta-\l}}(I_1(\eta,z)+I_2(\l,\eta,z)),\eqno(4.60)$$
where
$$\displaystyle \eqalign{
\bar I_1&=-\intt_{\C}{{\overline{ r(\zeta)}}\over {\zeta-\eta}}
e^{-iS(\zeta,z)}d\zeta_Rd\zeta_I,\cr
I_2&=\intt_{\C}{r(\zeta)\over (\zeta-\l)}{(\eta-\zeta)\over
(\bar\zeta-\bar\zeta)}
e^{iS(\zeta,z)}d\zeta_Rd\zeta_I,\cr}$$
where
$$iS(\zeta,z)=(-i/2)(\zeta\bar z +z/\zeta +\bar\zeta\bar z +
\bar z/\bar\zeta).
$$
Applying the formula (4.64) from the proof of lemma 7 we get
$$\eqalignno{
\bar I_1&={{\pi \overline{r(\eta)}e^{-iS(\eta,z)}}\over {(i/2)(z-
\bar z/\bar\eta^2)}}+\intt_{\C}{e^{-iS(\zeta,z)}\over {\zeta-\eta}}\\_
{\bar\zeta}\Bigl({{\overline{ r(\zeta)}}\over
{(i/2)(z-\bar z/\bar\zeta^2)}}\Bigr)d\zeta_Rd\zeta_I,\cr
I_2&={{\pi r(\l)e^{iS(\l,z)}}\over {(i/2)(z-\bar z/\bar\l^2)}}
\Bigl({{\eta-\l}\over {\bar\eta-\bar\l}}\Bigr)+\intt_{\C}{e^{iS(\zeta,z)}
\over {\zeta-\l}}\Bigl({{\eta-\zeta}\over {\bar\eta-\bar\zeta}}\Bigr)
\\_{\bar\zeta}\Bigl({r(\zeta)\over {(i/2)(z-\bar z/\bar\zeta^2)}}\Bigr)
d\zeta_Rd\zeta_I\cr
&+\intt_{\C}{e^{iS(\zeta,z)}\over {\zeta-\l}}{r(\zeta)
\over {(i/2)(z-\bar z/\bar\zeta^2)}}\\_{\bar\zeta}\Bigl({{\eta-\zeta}\over
{\bar\eta-\bar\zeta}}\Bigr)d\zeta_Rd\zeta_I=I_{20}+I_{21}+I_{22} .
&(4.61)\cr}$$
The following estimate is valid
$$|I_1|+|I_{20}|+|I_{21}|\le {{\rm const}\over |z|}, \eqno(4.62)$$
where the constant depends only on $b(\zeta)$.

It remains to estimate $I_{22}$.
$$\displaystyle I_{22}={1\over |z|}{1\over (\eta-\l)}\intt_{\C}
\Bigl({1\over {\zeta-\l}}-{1\over {\zeta-\eta}}\Bigr)
F(\zeta,\eta,z)d\zeta_Rd\zeta_I,$$
where
$$F(\zeta,\eta,z)=e^{iS(\zeta,z)}{{|z|r(\zeta)}\over
{(i/2)(z-\bar z/\bar\zeta^2)}}{(\eta-\zeta)^2\over (\bar\eta-\bar\zeta)^2}$$
and $F(\zeta,\eta,z)$ is a continuous rapidly decaying function of
$\zeta$ uniformly in $\eta,z$. From these properties of $F$ it follows
that
$$|I_{22}|<{{\rm const}\over |z|}{|\ln(|\eta-\l|)|\over
(1+|\eta-\l|\ln(|\eta-\l|))}, \eqno(6.63)$$
where the constant depends only on $b(\zeta)$.

The estimate (4.18) for $m=0,n=0$ follows from (4.60)-(4.63).

{\it The proof of lemma } 7.
Let us start from the following formula
$$\displaystyle \eqalign{ \\_{\bar\l}^{-1}\bigl(e^{iS(\l,z)}F(\l,z)\bigr)&=
{e^{iS}\over {iS_{\bar\l}}}\sum_{k=0}^{N-1}(-1)^k\Bigl(\\_{\bar\l}\circ
 {1\over {iS_{\bar\l}}}\Bigr)^kF(\l,z)\cr
&+(-1)^N\\_{\bar\l}^{-1}\Bigl[e^{iS}\bigl(\\_{\bar\l}\circ
{1\over {iS_{\bar\l}}}
\bigr)^NF(\l,z)\Bigr],\cr}\eqno(4.64)$$
where
$$\displaystyle \bigl(\\_{\bar\l}^{-1}f\bigr)(\l)=-{1\over \pi}\intt_{\C}
f(\zeta){{d\zeta_Rd\zeta_I}\over {\zeta-\l}}.\eqno(4.65)$$
This formula is formal, in general, however if all functions
$\bigl(\\_{\bar\l}
\circ {1\over {iS_{\bar\l}}}\bigr)^kF(\l,z),\ k=0,\ldots,N$ are
continuous in
$\l$ and vanishes as $\l\to\infty$ sufficiently fast then formula
(4.64) is an exact identity. Applying (4.64) to $A_z*f(z,\l)$ we get
$$\displaystyle \eqalign{
A_z*f(z,\l)&={{2i}\over (z-\bar z/\bar\l^2)}\exp \bigl[-{i\over 2}(\l\bar z +
\bar\l z +z/\l +\bar z/\bar\l)\bigr] \cr
&\times \Biggl[\sum_{k=0}^{N-1}(-1)^k\Bigl(\\_{\bar\l}\circ {{2i}\over
 {z-\bar z/\bar\l^2}}\Bigr)^k\Biggr]*(r(\l)\overline{ f(z,\l)})+R_N(z,\l),
 \cr}
\eqno(4.66)$$
$$\eqalign{
R_N(z,\l)&=(-1)^N\\_{\bar\l}^{-1}\Bigl[\exp\bigl[-{i\over 2}(\l\bar z +
\bar\l
 z+z/\l + \bar z/\bar\l)\bigr] \cr
&\times \Bigl(\\_{\bar\l}\circ {{2i}\over {z-\bar z/\bar
\l^2}}\Bigr)^N*(r(\l)\overline{ f(z,\l)})\Bigr]\cr
&={(-1)^N\over w^N}\\_{\bar\l}^{-1}\Bigl[\exp\bigl[-{i\over 2}(\l\bar z +
\bar\l
 z+z/\l+\bar z/\bar\l)\bigr] \cr
&\times \Bigl(\\_{\bar\l}\circ {1\over {1-\bar\nu/\bar\l^2}
}\Bigr)^N*(r(\l)\overline{ f(z,\l)})\Bigr].\cr}\eqno(4.67)$$
The function $\Bigl(\\_{\bar\l}\circ {1\over {1-\bar\nu/\bar\l^2}}\Bigr)^N*
(r(\l)\overline{ f(z,\l)})$ and all their derivatives are bounded on
the $\l$-plane
uniformly in $z$. Thus,
 $$\\_{\l}^{k_1}\\_{\bar\l}^{k_2}\\_z^{n_1}
\\_{\bar z}^{n_2}R_N(z,\l)=O\Bigl({1\over |z|^{N-k_1-k_2}}\Bigr),\ |z|\to
\infty \ {\rm for\ all}\ k_1,k_2,n_1,n_2\ge 0.\eqno(4.68)$$
It proves (4.31).Here we used the following property. Let $\varphi(z,\l)$
be an infinitely smooth function of $z,\l,\ z\ne 0$ and all the derivatives
$\\_z^{n_1}\\_{\bar z}^{n_2}\varphi(z,\l),\ n_1,n_2\ge 0$ are from the
Schwartz class in $\l$. Then
$$\\_z^{n_1}\\_{\bar z}^{n_2}\\_{\l}^{k_1}\\_{\bar\l}^{k_2}\\_{\bar\l}^{-1}*
\varphi (z,\l)=\\_{\bar\l}^{-1}\\_z^{n_1}\\_{\bar z}^{n_2}\\_{\l}^
{k_1}\\_{\bar\l}^{k_2}*\varphi (z,\l),\eqno(4.69)$$
where $\\_{\bar\l}^{-1}$ is defined by (4.65).

To calculate (4.32) consider (4.66) as $|\l|\to\infty$. We see, that
nontrivial
contribution arises only from the term $R_N(\l,z)$.
$$\displaystyle \eqalign{
 \alpha_{-k}(z)&={(-1)^N\over w^N}\Bigl({1\over \pi}\Bigr)
\intt_{\C}
\exp\bigl[-{i\over 2}(\l\bar z+\bar\l z+z/\l+\bar z/\bar\l)\bigr] \cr
&\times \Bigl(\\_{\bar\l}\circ {1\over {1-\bar\nu/\bar\l^2}}\Bigr)^N*
(r(\l)\overline{
f(z,\l)}\,\l^{k-1}d\l_Rd\l_I,\cr}\eqno(4.70)$$
$N$ can be chosen arbitrary large. It proves the statement 2.

Assume that $|\l|=1$. Then
$$A_z*f(z,\l)=R_N(z,\l) \eqno (4.71)$$
for any $N\ge 0$. It proves the statement 3.

Applying $A_z$ to the both sides of (4.66) we get
$$\displaystyle A_z^2*f(z,\l)=-\\_{\bar\l}^{-1}\Biggl\{r(\l){1\over {wR}}
\Biggl[\sum_{k=0}^{N-1}{1\over w^k}\Bigl(\\_{\l}\circ {1\over R}\Bigr)^k
\Biggr]
*(\overline{ r(\l)}f(\l,z))\Biggr\}+Q_N,\eqno(4.72)$$
where $w,R(\nu,\l)$ are defined by (4.34),
$$Q_N(z,\l)=\\_{\bar\l}^{-1}\Bigl\{r(\l)\exp\bigl[{i\over 2}(\l\bar z +
\bar\l z
+z/\l+\bar z/\bar\l\bigr]\overline{ R_N(z,\l)}\Bigr\},\eqno(4.73)$$
$R_N(z,\l)$ is defined by (4.67). Let $N>n_1+n_2$. Let us apply the
operator

$D=\\_z^{n_1}\\_{\bar z}^{n_2}\\_{\l}^{k_1}\\_{\bar\l}^{k_2}$ to (4.72).
Using
(4.68),(4.69) we get
$$\displaystyle \eqalign{
DA_z^2*f(z,\l)&=-\\_{\bar\l}^{-1}\Biggl\{D*\Bigl(\sum_{k=0}^{N-1}\Bigl[
{1\over R}{1\over w^{k+1}}\bigl(\\_{\l}\circ {1\over R}\bigr)^k\Bigr]*
(\overline{
 r(\l)} f(\l,z))\Bigr)\Biggr\}\cr
&+\\_{\bar\l}^{-1}\Bigl\{D*\Bigl(r(\l)\exp\bigl[{i\over 2}(\l\bar z+\l z+
z/\l+\bar z/\bar\l\bigr]\overline{ R_N(z,\l)}\Bigr)\Bigr\}.\cr}\eqno(4.74)$$
Taking into account (4.30) and the properties 1)-3) from theorem 3 we obtain
that the $k$- term,$\ 0\le k\le N-1$ in (4.74)  is $O\bigl({1\over |z|^{k+1}}
\bigr)$
as $|z|\to\infty $ uniformly in $\l$.  Combining it with (4.67) we complete
the
proof of the statement 4. Using that the number $N$ in (4.72),(4.74),(4.68)
can
be  taken  arbitrary we prove (4.33).
\medskip
{\it The proof of lemma } 9.
Let us observe that functions ${\cal C}_n(\l,\nu),\ n=0,\ldots,M$ satisfy
(4.39)
if and only if these functions satisfy the following system
$$\left\{\eqalign{
\\_{\bar\l}{\cal C}_m(\l,\nu)&={1\over R}r(\l) \overline{{\cal D}_m(\l,\nu)},
\ m=0,
\ldots,M \cr
{\cal D}_m(\l,\nu)&=\overline{\Bigl(\\_{\l}\circ {1\over R}\Bigr)}
{\cal D}_{m-1}(\l,\nu)
-r(\l) \overline{{\cal C}_{m-1}(\l,\nu)},\ m=1,\ldots,M, \cr}\right.
\eqno(4.75)$$
where
$${\cal D}_0(\l,\nu)=-r(\l),\eqno(4.76)$$
the functions ${\cal  C}_m(\l,\nu)$ are continuous in $\l$ for fixed $\nu$
and
$${\cal C}_m(\l,\nu)\to 0\ {\rm as}\ \l\to\infty.\eqno(4.77)$$
Let us observe also, that functions $c_m(\l)$ satisfy (4.48) if and only if
they
satisfy the following system
$$\left\{\eqalign{
\\_{\bar\l}c_m(\l)&=r(\l)\overline{d_m(\l)},\ \l\ne 0,\ m=0,\ldots,M \cr
d_m(\l)&=\\_{\bar\l}d_{m-1}(\l)-r(\l)\overline{c_{m-1}(\l)},\ \l\ne 0,\ m=1,
\ldots,M, \cr}\right.\eqno(4.78)$$
where
$$d_0(\l)=-r(\l).\eqno(4.79)$$
The system (4.78) coincides with (3.5) but  with a different starting
function
$d_0(\l)$ instead of $b_0(\l)$.

Let us prove that functions $c_m(\l)$ from  (4.49),$m=0,\ldots,M$ and
functions
$$\displaystyle d_m(\l)=-\pi\,{\rm sign}\,(\l\bar\l-1)\sum_{k=0}^m\overline{
\alpha_{mk}(\l)}\ b_k(\l)\eqno(4.80)$$
solves (4.78). The direct calculation with help of (3.5) shows that

\medskip

$$\displaystyle \eqalign{
\pls c_m(\l)=&\sg\sum_{k=0}^m\pls\, (\a_{mk}(\l)a_k(\l)) \cr
&=\sg\s k0m{\a}{mk}{\l}\pls\, a_k(\l) \cr
&=\sg\s k0m{\a}{mk}{\l} r(\l)\overline{b_k(\l)}\cr
&=r(\l)\Bigl(\sg\,\overline{\sum_{k=0}^m\overline{\alpha_{mk}(\l)} b_k(\l)}
\Bigr)=r(\l)
\overline{d_m(\l)},\cr}\eqno(4.81)$$
$$\displaystyle \eqalign{
&\pls\,d_m(\l)=\sg\,\sum_{k=0}^m\pls\,(\overline{\a_{mk}(\l)}\,b_k(\l))=\sg
\cr
&\times \Biggl[\sum_{k=0}^m(\overline{\pls\,\a_{mk}(\l)})\,b_k(\l)+
\sum_{k=0}^m\overline{\a_{mk}(\l)}\,(b_{k+1}(\l)+r(\l)\,\overline{a_k(\l)})\
Biggr]\cr
&=\sg\,\Biggl[\sum_{k=0}^m\overline{\a_{m+1,k}(\l)}\,b_k(\l)-\sum_{k=1}^m
\overline{\a_{m,k-1}(\l)}\,
b_k(\l)\cr
&+\sum_{k=0}^m\overline{\a_{mk}(\l)}\,b_{k+1}(\l)
+r(\l)\sum_{k=0}^m\overline{\a_{mk}(\l)}\,\overline{a_k(\l)}\Biggr] \cr
&=\sg\,\sum_{k=0}^{m+1}\overline{\a_{m+1,k}(\l)}\,b_k(\l)\cr
&+r(\l)\overline{
\Biggl(\sg\,\sum_{k=0}^m\a_{mk}(\l)a_k(\l)\Biggr)}=d_{m+1}(\l)+r(\l)
\overline{c_m(\l)}.\cr}\eqno(4.82)$$
In these  calculations and later we use the fact that the functions
$a_k(\l),b_k(\l),\ k=0,\ldots,M$\ vanishes on the unit circle $|\l|=1$
with all derivatives. (This property follows from conditions 1),3),4) of
 theorem 1.) Due to this property the functions $c_n(\l)$ defined by (4.49)
and $d_n(\l)$ defined by (4.80) are smooth in the neighborhood of the unit
circle $|\l|=1$.

Let us prove now that functions ${\cal C}_n(\l,\nu)$ defined by (4.47) and
${\cal D}_n(\l,\nu)$ defined by
$$\displaystyle
{\cal D}_n(\l,\nu)=\bar R\,\sum_{l=0}^n\overline{f_{nl}(\l,\nu)}\,d_l(\l)
\eqno(4.83)$$
satisfy (4.75).

The direct calculation with help of (4.78),(4.45) shows that
$$\displaystyle\eqalign{
&\pls\,{\cal C}_m(\l,\nu)=\pls\,\s k0mf{mk}{\l,\nu}\,c_m(\l)=
\s k0mf{mk}{\l,\nu}\,\pls\,c_m(\l)\cr
&=\s k0mf{mk}{\l,\nu}\,r(\l)\,\overline{d_k(\l)}=
r(\l)\,{1\over R(\l,\nu)} \cr
&\times \overline{
\Biggl(\sum_{k=0}^m\overline{R(\l,\nu)}\,\overline{f_{mk}(\l,\nu)}d_k(\l)
\Biggr)}
={r(\l)\over R(\l,\nu)}\,\overline{{\cal D}_m(\l,\nu)},\cr}\eqno(4.84)$$
$$\displaystyle\eqalign{
&\overline{\Bigl(\pl \circ {1\over R(\l,\nu)}\Bigr)}\,{\cal D}_m(\l,\nu)=
\pls\,\sum_{l=0}^m\overline{f_{nl}(\l,\nu)}\,d_l(\l)\cr
&=\sum_{l=0}^m\overline{R(\l,\nu)}\,\overline{f_{n+1,l}(\l,\nu)}\,d_l(\l)-
\sum_{l=1}^m\overline{f_{n,l-1}(\l,\nu)}\,d_l(\l)+\sum_{l=0}^m
\overline{f_{nl}(\l,\nu)}\,d_{l+1}(\l)\cr
&+\sum_{l=o}^m\overline{f_{nl}(\l,\nu)}\,r(\l)\,\overline{c_l(\l)}=
\sum_{l=0}^{m+1}\overline{R(\l,\nu)}\,
\overline{f_{n+1,l}(\l,\nu)}\,d_l(\l) \cr
&+r(\l)\,\Biggl(\overline{
\s l0mf{nl}{\l,\nu}\,c_l(\l)}\Biggr)
={\cal D}_{m+1}(\l,\nu)+r(\l)\,\overline{{\cal C}_m(\l,\nu)}.\cr}
\eqno(4.85)$$
Using the estimates (4.44),(4.50),(3.4) it is easy to show that the
function
${\cal C}_m(\l,\nu),$
\noindent
${\cal D}_m(\l,\nu),\ m=0,\ldots,M$\ are bounded in $\l$ and
${\cal C}_m(\l,\nu)\to 0$ as $\l\to\infty$.

Lemma 9 is proved.
\bigskip
\noindent
{\bf 5.Two uniqueness theorems.}
\medskip
\noindent
{\it Definition}: a measurable potential $v(z)$ will be called exponentially
decreasing if there exist $\alpha>0$ and $\beta>0$ such that
$|v(x)|<\beta e^{-\alpha |x|}$.
\medskip\noindent
{\bf Theorem 3.}
{\it Let the fixed energy scattering   amplitude of two exponentially
decreasing
potentials with the property } (0.2) {\it coincide and one of these
potentials
possesses, in addition, the "small norm" property } (0.5) {\it at  this
fixed energy.
Then these two   potentials coincide.}
\medskip
\noindent
{\bf Corollary 3.}
\noindent
{\it There exist no nonzero two-dimensional exponentially decreasing real
 nonsingular potentials transparent at a fixed energy.} ({\it There is no
 "small norm" assumption in corollary} 3).
\medskip
\noindent
{\it Remark.}
The result of theorem 1 improve the corresponding result from [9,\ 12]. The
proof uses, in particular, the ideas from [27].
\medskip
\noindent
{\it The proof of theorem } 3. We shall use (see [12,\ 27]) the fact that
for an
exponentially decreasing potential with property (0.2) each of functions
$a(\l),b(\l)$ in the domains $D_{+}$ and $D_{-}$  can be written as a ratio
of two real analytic functions; the Fredholm determinant $\Delta(\l)$ of the
equation (1.3) is real analytic in $D_{+}$ and $D_{-}$ and all these three
functions are uniquely determined by the scattering amplitude at fixed energy.
One of the potentials satisfies the "small norm" assumption (0.5), thus
$\Delta(\l)\ne 0$ for all $\l\in D_{+},D_{-}$.

Thus for both potentials $v_1(z),v_2(z)$ the corresponding functions
$\mu_1(z,\l),\mu_2(z,\l)$ satisfy equations $(1.19^{\prime}),
(1.20^{\prime})$,
where $\rho(\l,\lp)$ is defined by (1.28). But one of the potentials
satisfies
the "small norm" assumption (0.5) and it is shown in [12] that equations
$(1.19^{\prime}),(1.20^{\prime})$ have unique solution, i.e. $\mu_1(z,\l)=
\mu_2(z,\l)$. Thus, $v_1(z)=v_2(z).$
\medskip
\noindent
{\bf Theorem 4.}
{\it Let the potential } $v(z)$ {\it satisfy } (0.2) {\it and its forward
scattering amplitude}
$f(k,k)$ {\it is identically zero at an energy interval\ } ${\rm E_{fix}}\
-\delta<k^2<
\ {\rm E_{fix}}\ +\delta$. {\it Then the potential } $v(z)$ {\it is equal
to zero
identically} . (In this theorem we do not use the "small norm" assumption).
This theorem and its proof given below are valid in   any dimension \
${\rm dim}\ =1,2,3\ldots.$

{\it The proof of theorem } 4. As a consequence of the unitarity property
(1.29)
of the scattering operator we have the well known "optical theorem"
$$\displaystyle {\rm Im}\ f(k,k)=-{\pi\over 2\sqrt{E}}\int_{l^2=k^2}
f(k,l)\overline{f(k,l)}dl. \eqno(5.2)$$
{}From (5.2) it follows that ${\rm Im}\ f(k,k)=0$ if and only if $f(k,l)=0$
for all $l$ such that $l^2=k^2$. So, if at the energy level $E$ the forward
scattering amplitude $f(k,k)=0,\ k^2=E$ then the whole fixed-energy
scattering amplitude $f(k,l),\ k^2=l^2=E$ is equal to zero. It is well known
that the forward scattering amplitude $f(s\gamma,s\gamma),\ \gamma\in \R^2,
|\gamma|=1,\ s\in\R_+$ admits a meromorphic continuation in $s$\ to the upper
half  plane. Thus, if the forward
scattering amplitude is equal to zero on an energy interval it is equal to
 zero for all energies. Thus, the whole scattering amplitude at all energies
is
equal to zero and as a consequence the potential $v(z)$ is identically zero.
\medskip\noindent
{\it Remark.}
The result of theorem 2 is valid also for the  equation
$$-\Delta\psi-k^2u(x)\psi=k^2\psi,\ x\in \R^d,\ d=1,2,3\ldots,\eqno(5.3)$$
where $u(x)$ is a real measurable function such that $|u(x)| < q/(1+|x|)
^{d+\varepsilon}$ and for the equation
$$-\Delta\psi+(v(x)-k^2u(x))\psi=k^2\psi,\ x\in \R^d,\ d=2,3,4,\ldots,
\eqno(5.4)$$
where $v(x),u(x)$ are real measurable functions such that
$$|v(x)|<{q_1\over (1+|x|)^{d+\varepsilon}},\ |u(x)|<{q_2\over (1+|x|)^
{d+\varepsilon}}.$$
For equation (5.4) the result that both potentials $v(x)$ and $u(x)$ are
equal
identically to zero if the scattering amplitude $f(k,l)=0$ for all
$k,l\in \R^d,\ d\ge 2,k^2=l^2$ is a corollary of results obtained in [25].
\bigskip\noindent
{\bf 6.Nonlinear integrable equations.}
\medskip\noindent
In this section we discuss if the additional conditions on the scattering
data
studied in sections 3,4 are invariant under deformations, generated by
nonlinear
 equation (0.8) and its higher analogs .

In terms of the scattering data these equations (Novikov-Veselov equations)
take the form
$$\eqalign{
\ch b{}{} &=i\bigl(\l^{2l+1}+{1\over \l^{2l+1}} +\bar\l^{2l+1}+
{1\over \bar\l^{2l+1}}\bigr)b(\l,t),\cr
\ch f{}{\lp} &=i\bigl(\l^{2l+1}+{1\over \l^{2l+1}}
 -(\lp)^{2l+1}-\o \bigr)f(\l,\lp,t),\cr
\ch {\rho}{}{\lp} &=i\bigl(\l^{2l+1}+{1\over \l^{2l+1}}
 -(\lp)^{2l+1}-\o \bigr)\rho(\l,\lp,t),\cr
\ch h{\pm}{\lp} &=i\bigl(\l^{2l+1}+{1\over \l^{2l+1}}
 -(\lp)^{2l+1}-\o \bigr)h_{\pm}
(\l,\lp,t)\cr}\eqno(6.1)$$
(equation (0.8) corresponds to $l=1$).

\medskip

It is known that the symmetry conditions (1.25),(1.29),(1.24) (and, as a
corollary,  (3.20)) are invariant under the flows (6.1). For the additional
conditions from section 3 the situation is more interesting.
\medskip
\noindent
{\bf Theorem 5.}

\item{1)} {\it Let the scattering data } $b(\l,t),f(\l,\lp,t)$ {\it satisfy }
(6.1), {\it where at} $t=0$
$b(\l,0)\in C_3^{(\infty)}(D_-)$,
\noindent
$f(\l,\lp,0)\in C^{(\infty)}(T^2)$
 {\it and } $b(\l,0),f(\l,\lp,0)$ {\it satisfy } (1.25),(1.29) {\it and
 the first two additional
 conditions}  (3.11),(3.12) {\it from section } 3 {\it corresponding to }
 $M=0$. {\it Then these
conditions are fulfilled for all } $t$.
\item{2)} {\it Let the scattering data } $b(\l,t),f(\l,\lp,t)$ {\it satisfy}
(6.1) {\it with}
$l=1$, {\it where at}

\noindent
$t=0,
b(\l,0)\in
C_3^{(\infty)}(D_-),f(\l,\lp,0)\in C^{(\infty)}(T^2)$ {\it and} $b(\l,0),
f(\l,\lp,0)$ {\it satisfy} (1.25),

\noindent
(1.29) {\it and the first 4 additional conditions} (3.11),(3.12) {\it and}
(3.15),(3.16) {\it for } $n=0$ ({\it corresponding to} $M=1$).
{\it Then these conditions are fulfilled for all } $t$ {\it if and only if }
$a_0^-(\infty)
=0$. (Let us recall that for the potential $v(z)$ with the property (0.2)
$a_0^-(\infty)=\hat v(0)$, where $\hat v(p)$ is the Fourier transform
(3.34) of $v(z)$).

{\it The proof of theorem } 5.

Let $|\l|=1,\ \lp=-\l$. Then
$$ b_0^-(\l,t)=e^{2i(\l^{2l+1}+\bar\l^{2l+1})t}\,b_0^-(\l,0), \eqno(6.2)$$
$$h_-(\l,\lp,t)=e^{2i(\l^{2l+1}+\bar\l^{2l+1})t}h_-(\l,\lp,0)\eqno(6.3)$$
and (3.11) is fulfilled identically for all $t$ if it is fulfilled for $t=0$.

We have
$$\eqalign{
h_-(\l,\lp,t)&=h_-(\l,\lp,0)\ {\rm for}\ |\l|=1,\lp=\l,\cr
b(\l,t)\,\overline{b(\l,t)}&=b(\l,0)\,
\overline{b(\l,0)}\ {\rm for}\ \l\in D_- \cr}\eqno(6.4)$$
and (3.12) does not depend on $t$ and it is fulfilled identically for all $t$
if it is fulfilled for $t=0$. (The first part of theorem 5 is proved).

Thus, for $a_0^-(\l,t)$ defined by (3.13) we have
$$a_0^-(\l,t)=a_0^-(\l,0).\eqno(6.5)$$
{}From (3.13) using the symmetries $h_-(\l,\l)=h_-(-\l,-\l)$ (it is a
consequence
of (1.30)) and (1.25) we get that
$$a_0^-(-\l,t)=a_0^-(\l,t).\eqno(6.6)$$
{}From (3.14) it follows that
$$\eqalign{
&b_1^-(\l,t)\big|_{|\l|=1}=\pls\, b_0^-(\l,t)-{\pi\over \bar\l}b_0^-(\l,t)\,
\overline{a_0^-(\l,t)}\big|_{|\l|=1}\cr
&=e^{2i(\l^{2l+1}+\bar\l^{2l+1})t}b_1^-(\l,0)+(2l+1)it(\bar\l^{2l}-\l^
{2l+2})b_0^-(\l,t).\cr}\eqno(6.7)$$
$$\eqalign{
&(-i\lp\\_{\varphi^{\prime}})h_-(\l,\lp,t)\bigg|_
{\scriptstyle \lp=-\l\atop\scriptstyle |\l|=1}={\lp}^2\\_{\lp}h_-(\l,\lp,t)
\bigg|_
{\scriptstyle \lp=-\l\atop\scriptstyle |\l|=1}\cr
&=e^{2i(\l^{2l+1}+\bar\l^{2l+1})t}{\lp}^2\\_{\lp}h_-(\l,\lp,0)
\big|_{\lp=-\l}\cr
&+(2l+1)it(\bar\l^{2l}-\l^{2l+2})h_-(\l,\lp,t)\big|_{\lp=-\l}.\cr}
\eqno(6.8)$$

Comparing (6.7) and (6.8) and using the fact that (3.11) is   fulfilled for
all $t$ we get that if under conditions of the first part of theorem 5\ \ \
(3.15) with $n=0$ is fulfilled for $t=0$ then it  is
fulfilled for all $t$.

{}From (3.5),(6.1),(6.5) it follows that
$$\eqalign{
&{\pi\over \bar\l}b_0^-(\l,t)\overline{b_1^-(\l,t)}
={\pi\over \bar\l}b_0^-(\l,t)\pl \overline{b_0^-(\l,t)}-
{\pi^2\over \l\bar\l}b_0^-(\l,t)\,\overline{b_0^-(\l,t)}\cr
&\times a_0^-(\l,t)=-(2l+1)it\bigl(\l^{2l}-{1\over \l^{2l+2}}\bigr)b_0(\l,0)
\,\overline{b_0(\l,0)}+\pls\, a_1^-(\l,0)\cr
&=\pls\,  a_1^-(\l,0)-(2l+1)it\pls\, A_1^-(\l),\cr}\eqno(6.9)$$
where
$$\pls\, A_1^-(\l)=\bigl(\l^{2l}-{1\over \l^{2l+2}}\bigr)\pls\, a_0^-(\l,0).
\eqno(6.10)$$
{}From (6.1) it follows that
$$
\\_{\lp}\,h_-(\l,\lp,t)\bigg|_{\scriptstyle \lp=\l\atop\scriptstyle
|\l|=1}=\\_{\lp}h_-(\l,\lp,0)
-(2l+1){\rm it}\bigl(\l^{2l}-{1\over \l^{2l+2}}\bigr)h_-(\l,\lp,0)
\bigg|_{\scriptstyle
\lp=\l\atop\scriptstyle |\l|=1}.\eqno(6.11)$$
Let $l=1$. Using (6.9)--(6.11) we can transform the boundary value
problem (3.5a),

\noindent
(3.4a-),(3.9a-) with  $m=1$ to the following form
$$\eqalignno{
\pls\, A_1^-(\l)&=\bigl(\l^2-{1\over \l^4}\bigr)\,\pls\,a_0^-(\l,0),\ \l\in
D_-, &(6.12)\cr
A_1^-(\l)&=\bigl(\l^2-{1\over \l^4}\bigr)\,h_-(\l,\lp,0)\bigg|_{\scriptstyle
|\l|=1\atop\scriptstyle \lp=\l}, &(6.13)\cr
A_1^-(\l)&=O(1),\ \l\to\infty. &(6.14)\cr}$$
If $b_1^-(\l,t)$ is  expressed via $b(\l,t),h_-(\l,\lp,t)$ then this boundary
value problem is  equivalent to (3.16),$n=0$.

{}From (6.12),(6.14),(6.6) it follows that
$$A_1^-(\l)=\bigl(\l^2-{1\over \l^4}\bigr)\,a_0^-(\l)-\l^2\,a_0^-(\infty)+
\varphi(\l) \eqno(6.15)$$
for some $\varphi(\l)$ such that $\varphi(\l)$ is a bounded holomorphic
function on $D$.

Substituting (6.15) to (6.13) and using (3.4a) with $m=0$ we get
$$\bigl(\l^2-{1\over \l^4}\bigr)\,a_0^-(\infty)+\varphi(\l)\big|_{|\l|=1}
=0.\eqno(6.16)$$

The problem of finding a bounded holomorphic function $\varphi(\l)$ on
$D_-$ with the boundary value (6.16) is solvable if and only if
$a_0^-(\infty)=0$. Under these conditions $\varphi(\l)=0$.
So, under conditions of the second part of theorem 5 the 4-th
condition is fulfilled identically in $t$ if and only if the first 4
conditions are fulfilled for $t=0$ and $a_0^-(\infty)=0$.

Theorem 5 is proved.
\vfill\eject
{\bf References}
\medskip
{\ninerm
\item{ 1.} Manakov,\ S.V.:\ The inverse scattering method and two-dimensional
 evolution equations.\ Uspekhi Mat.\ Nauk\ {\bf 31}(5),\ 245-246\ (1976)\
(in russian)
\item{ 2.} Dubrovin,\ B.A.,\ Krichever,\ I.M.,\ Novikov,\ S.P.:\ The
Schr\"odinger equation in a periodic field and Riemann surfaces.\ Dokl.
\ Akad.\ Nauk
\ SSSR\ {\bf 229},\ 15-18\ (1976),\ translation in Sov.\ Math.\ Dokl.\
{\bf 17}\ ,
947-951\ (1976)
\item{ 3.} Veselov,\ A.P.,\ Novikov,\ S.P.:\ Finite-zone, two-dimensional,
potential Schr\"odinger operators. Explicit formulas and evolution equations.
\ Dokl.\ Akad.\ Nauk\ SSSR\ {\bf 279},\ 20-24\ (1984),\ translation in Sov.\
Math.\ Dokl.\ {\bf 30},\ 588-591\ (1984)
\item{ 4.} Veselov,\ A.P.,\ Novikov,\ S.P.:\ Finite-zone, two-dimensional
Schr\"odinger operators.\ Potential operators.\ Dokl.\ Akad.\ Nauk.\ SSSR\
{\bf 279},\ 784-788\ (1984),\ translation in Sov.\ Math.\ Dokl.\ {\bf 30},
\ 705-708\ (1984)
\item{ 5.} Grinevich,\ P.G.,\ Novikov,\ R.G.:\ Analogues of multisoliton
potentials for the two-dimensional Schr\"odinger operator.\ Funkt.\ Anal.\
i Pril.\ {\bf 19}(4),\ 32-42\ (1985)\ translation in Funct.
\ Anal. and Appl.\
{\bf 19},\ 276-285\ (1985)
\item{ 6.} Grinevich,\ P.G.,\ Novikov,\ R.G.:\ Analogues of multisolition
potentials for the two-dimensional Schr\"odinger equations and a nonlocal
 Riemann problem.\ Dokl.\ Akad.\ Nauk SSSR {\bf 286},\ 19-22\ (1986),\
translation in Sov.\ Math.\ Dokl.\ {\bf 33},\ 9-12\ (1986)
\item{ 7.} Novikov,\ R.G.:\ Construction of a two-dimensional Schr\"odinger
operator with a given scattering amplitude at fixed energy.\ Teoret.\ Mat.\
Fiz.\ {\bf 66}\ 234-240\ (1986)
\item{ 8.} Grinevich,\ P.G.,\ Manakov,\ S.V.:\ The inverse scattering problem
for the two-dimensional  Schr\"odinger operator,\ the $\bar\\-$ method and
non-linear equations.\ Funkt.\ Anal.\ i Pril.\ {\bf 20}(2),\ 14-24\ (1986),\
translation in Funkt.\ Anal.\ and Appl.\ {\bf 20},\ 94-103\ (1986)
\item{ 9.} Novikov,\ R.G.:\ Reconstruction of a two-dimensional Schr\"odinger
operator from the scattering amplitude at fixed energy.\ Funkt.\ Anal.\ i
Pril.\ {\bf 20}(3),\ 90-91\ (1986),\ translation in Funkt.\ Anal.\ and Appl.\
{\bf 20},\ 246-248\ (1986)
\item{10.} Grinevich,\ P.G.:\ Rational solutions of the Veselov-Novikov
equations\ --\ Two-dimensional potentials that are reflectionless for fixed
energy.\ Teoret.\ Mat.\ Fiz.\ {\bf 69}(2),\ 307-310\ (1986)
\item{11.} Grinevich,\ P.G.,\ Novikov,\ S.P.:\ Two-dimensional "inverse
scattering problem" for negative energies and generalized-analytic functions.
\ I.\ Energies below the ground state.\ Funkt.\ Anal.\ i Pril.\ {\bf 22}(1),\
23-33\ (1988),\ translation in Funkt.\ Anal.\ and Appl.\ {\bf 22},\ 19-27\
(1988)
\item{12.} Novikov,\ R.G.:\ The  inverse scattering problem on a fixed energy
level for the two-dimensional Schr\"odinger operator.\ J.\ Funkt.\ Anal.\
{\bf 103},\ 409-463\ (1992)
\item{13.} Fran\c coise,\ J.-P.,\ Novikov,\ R.G.:\ Solutions rationnelles
des \'equations de type Korteweg-de Vries en dimension $2+1$ et probl\`emes
\`a $m$ corps sur la droite.\ C.R.\ Acad.\ Sci.\ Paris\ {\bf 314},\ S\'er.
I,\ 109-113\ (1992)
\item{14.} Boiti,\ M.,\ Leon,\ J.,\ Manna,\ M.,\ Pempinelli,\ F.:\ On a
spectral transform of a KDV-like equation related to the Schr\"dinger
operator in the plane.\ Inverse Problems\ {\bf 3},\ 25-36\ (1987)
\item{15.} Tsai,\ T.Y.:\ The Schr\"odinger operator in the plane.\
Dissertation,\ Yale University\ (1989)
\item{16.} Calderon,\ A.P.:\ On an inverse boundary value problem.\ Seminar
on Numerical Analysis and its Applications to Continuum Physics,\ Soc.\
Brasileira de Mathem\`atica,\ Rio de Janeiro,\  65-73\ (1980)
\item{17.} Kohn,\ R.,\ Vogelius,\ M.:\ Determining conductivity by boundary
measurements II.\ Interior results.\ Comm.\ Pure Appl.\ Math.\ {\bf 38},\
644-667\ (1985)
\item{18.} Sylvester,\ J.,\ Uhlmann,\ G.:\  A uniqueness theorem for an
inverse boundary value problem in electrical prospection.\ Comm.\ Pure Appl.\
Math.\ {\bf 39},\ 91-112\ (1986)
\item{19.}  Novikov,\ R.G.:\ Multidimensional inverse spectral problem for the
 equation  $-\Delta\psi + (v(x)-Eu(x))\psi=0$.\ Funkt.\ Anal.\ i Pril.\
{\bf 22}
(4),\ 11-22\ (1988),\ translation in Funct.\ Anal.\ and Appl.\ {\bf 22},\
263-272\ (1988)
\item{20.} Sun,\ Z.,\ Uhlmann,\ G.:\ Generic uniqueness for an inverse boundary
value  problem.\ Duke Math.\ J.\ {\bf 62},\ 131-155\ (1991)
\item{21.} Sun,\ Z.,\ Uhlmann,\ G.:\ Recovery of singularities for  formally
determined inverse problems.\ Commun.\ Math.\ Phys.\ {\bf 153},\ 431-445\
(1993)
\item{22.} Nachman,\ A.I.:\  Global uniqueness for a two-dimensional inverse
boundary value problem.\ Preprint\ (1993)
\item{23.} Faddeev,\ L.D.:\ Inverse problem of quantum scattering theory II.\
Itogi Nauki i Tekhniki,\ Sov.\ Prob.\ Mat.\ {\bf 3},\ 93-180\ (1974),\
translation in J.\ Sov.\ Math.\ {\bf 5},\ 334-396\ (1976)
\item{24.} Novikov,\ R.G.,\ Henkin,\ G.M.:\ The $\bar\\-$ equation in the
multidimensional inverse scattering problem.\ a)Preprint 27M,\ Institute
of Physics,\ Krasnoyarsk (1986),\ (in russian);\ b)Uspekhi Mat.\ Nauk\
{\bf 42}(3),\ 93-152\ (1987),\ translation in Russ.\ Math.\ Surv.\ {\bf 42}
(4),\ 109-180\ (1987)
\item{25.} Novikov,\ R.G.:\ Multidimensional inverse scattering problem
for acoustic equation.\ Preprint 32M,\ Institute of Physics,\ Krasnojarsk\
(1986) (in russian)
\item{26.} Henkin,\ G.M.,\ Novikov,\ R.G.:\ A multidimensional inverse
problem in  quantum and acoustic scattering.\ Inverse Problems\ {\bf 4},\
103-121\  (1988)
\item{27.} Novikov,\ R.G.:\ The inverse scattering problem at fixed
 energy for the three-dimensional Schr\"odinger equation with an exponentially
decreasing potential.\ Commun.\ Math.\ Phys.\ {\bf 161},\ 569-595\ (1994)
\item{28.} Manakov,\ S.V.:\ The inverse scattering transform for the time
dependent Schr\"odinger equation and Kadomtsev-Petviashvili equation.\
Physica D\ {\bf 3}(1,2),\ 420-427\ (1981)
\item{29.} Ablowitz,\ M.J.,\ Bar Yaacov,\ D.,\ Fokas,\ A.S.:\ On the inverse
scattering transform for the Kadomtsev-Petviashvili equation.\ Studies in
Appl.\ Math.\ {\bf 69},\ 135-143\ (1983)
\item{30.} Zakharov,\ V.E.:\ Shock waves spreading along solitons on the
surface of liquid.\ Izv.\ Vuzov Radiofiz.\ {\bf 2969},\ 1073-1079\ (1986)
\item{31.} Regge,\ T.:\ Introduction to complexe orbital moments.
Nuovo Cimento. \ {\bf 14},\ 951-976\ (1959)
\item{32.} Newton,\  R.G.:\ Construction of potentials from the phase
shifts at fixed energy.\ J.\ Math.\ Phys.\  {\bf 3},\ 75-82\ (1962)
\item{33.} Sabatier,\ P.C.:\ Asymptotic properties of the potentials in the
inverse--scattering problem at  fixed energy.\ J.\ Math.\ Phys.\ {\bf 7},
\ 1515-1531\ (1966)
\item{34.} Vekua,\ I.N.:\ Generalized analytic functions.\ Pergamon Press,\
Oxford 1962
\item{35.} Landau,\ L.D.,\ Lifshitz,\ E.M.:\ Quantum mechanics:\ Non-
relativistic theory.\ 2-nd ed.\ Pergamon Press,\ Oxford 1965
\end